THE RISK–RETURN RELATION IN THE CORPORATE LOAN MARKET

M. A. Duran,[1]

University of Malaga (Spain)

**Abstract:** This paper analyzes the hypothesis that returns play a risk-compensating role in the market for corporate revolving lines of credit. Specifically, we test whether borrower risk and the expected return on these debt instruments are positively related. Our main findings support this prediction, in contrast to the only previous work that examined this problem two decades ago. Nevertheless, we find evidence of mispricing regarding the risk of deteriorating firms using their facilities more intensively and during the subprime crisis.

**J.E.L. Codes:** G29, G39

**1. Introduction**

Revolving credit agreements play a major role in corporate financing. Ergungor (2001) points out that they represent around 80% of all commercial and industrial lending by US banks, and Sufi (2009) documents that a similar percentage of public firms have at least one of these credit instruments. According to the Board of Governors of the Federal Reserve System, drawdowns on syndicated credit lines represent around half of total loans outstanding in the US syndicated loan market, with the unused portion of revolving facilities far exceeding the used portion.[2]

---

[1] I thank Charles Calomiris and Anthony Saunders for helpful comments and suggestions. I also acknowledge financial support from the Spanish Ministry of Economics and Competitiveness (Project RTI2018-097620-B-I00), the European Regional Development Fund (Project FEDERJA-169), and the Salvador de Madariaga Program.
[2] See https://www.federalreserve.gov/releases/efa/efa-project-syndicated-loan-portfolios-of-financial-institutions.htm.

The primary rationale that economic modeling has identified for this widespread use of revolving facilities is that they provide firms insurance against tougher conditions in accessing credit (Campbell 1978, Holmström and Tirole 1988). The protection provided by this financing instrument stems from the right to borrow repeatedly at predetermined terms. As long as the facility is open and its credit limit is not reached, the borrower can exert this right when drawing on the revolving line involves less onerous conditions than borrowing in the market, due, for instance, to a loss in corporate creditworthiness. If the firm uses this right, it forces the lender to share in any decline in firm value, with the corresponding rent transferring from the lender to the borrower (Hawkins 1982). Thus, although credit risk is inherent in any lending relationship, it is qualitatively different in the market for revolving facilities: while a borrower's quality deteriorates, the borrower can still exercise the right to draw down according to a contractually set price scheme. Moreover, the lender cannot be certain about whether, when, or how much the borrower will draw down (Ho and Saunders 1983).

To avoid potential rent transfers between lenders and borrowers, credit line agreements include a large set of covenants, for instance, those allowing lenders to decline to make additional loans or even to accelerate the facility (Wight et al. 2007). These covenants limit the risk that borrowers can take on and hence constrain the menu of feasible risk–return combinations. However, before covenants are triggered, there is room for returns to compensate lenders for increasing corporate risk (Asquith et al. 2005). Indeed, if the return adequately fulfills this risk-compensating role, redistributions of wealth from the lender to the borrower could be prevented (Chava and Roberts 2008, n. 9).

The aim of this paper is to examine empirically this potential compensating role of the returns of revolving lines. The empirical literature has analyzed the risk–return nexus in financial intermediation mainly based on studying how aspects such as internationalization strategies, product mix, or funding decisions affect banks' yield or risk (DeYoung and Roland 2001, Stiroh 2004, Stiroh and Rumble 2006, Lepetit et al. 2008, Berger et al. 2010, Demirgüç-Kunt and Huizinga 2010, Maudos 2017). Our paper adopts a different perspective. We analyze directly the relation between risk and return in the market for credit lines, aiming at testing whether returns on this credit instrument carry a risk premium.

Regarding the empirical literature on revolving facilities, Asarnow and Marker (1995) study the risk–return relation in this market. Nevertheless, despite the relevance that the risk-compensating role of returns can have in diminishing potential lender–borrower conflicts (Jensen and Meckling 1976), no other empirical work on credit lines has delved into this problem in the last two decades. This lack of research is even more concerning not only because of the importance of credit facilities in corporate financing (Ergungor 2002, Kashyap et al. 2002, Jimenez et al. 2009, Sufi 2009), but also because Asarnow and Marker (1995) find evidence of mispricing.

Works closely related to ours are those that examine the relation between borrowers' creditworthiness and pricing provisions included in credit line contracts (Shockley and Thakor 1997, Strahan 1999, Beatty et al. 2002, Asquith et al. 2005, Manso et al. 2010, Berg et al. 2016, Duran 2017). Nevertheless, study of the risk-compensating role of returns is related to but differs substantially from the analysis of the relation between corporate risk and those provisions. Since outstanding drawdowns are not constant, returns depend not only on the charges to the borrower, but also on usage. Indeed, as Jones and Wu (2015) point out, lender profitability does not necessarily increase with interest rates and fees: higher costs for borrowers can decrease usage enough to reduce returns. Additionally, to fully grasp whether credit line agreements adequately charge for risk, the analysis should be based on data reflecting time variations in returns and risk. However, previous research on the relation between borrower risk and facility contractual pricing focuses on interest rates and fees at origination.

From a theoretical perspective, earlier research has also proposed using option pricing theory to value revolving facilities (Bartter and Rendleman 1979, Thakor et al. 1981, Hawkins 1982, Thakor 1982, Ho and Saunders 1983, Chateau 1990). This approach compares credit lines to put options, describing them as providing the right to sell a security (firm debt) at a prespecified price. Nevertheless, the modeling of credit lines as put options omits consideration of the basic features of this lending instrument that could be key to empirically examining whether returns compensate for increasing risk. In this regard, common simplifying assumptions in the literature are that the available borrowing amount can only be used in full or not at all, the right to drawdown can only be exercised at maturity, the interest rate and fee structure is substantially simplified, and a large proportion of the market is ignored because of the concrete type of credit line modeled (Ergungor 2001).

To study whether credit lines charge adequately for risk, this paper directly focuses on whether there is a trade-off between risk and returns. In particular, we use expected returns. The analysis in terms of this type of yield is consistent with the empirical financial intermediation literature, which suggests that lenders are risk averse (Ratti 1980, Hughes et al. 1995, 1996, 1999, Hughes and Moon 1997, Angelini 2000). Thus, the specific empirical hypothesis that we test is whether risk and expected returns are positively related.

A possible reason for the scarcity of research on this hypothesis is that it is highly demanding in terms of data, requiring knowledge of how usage, interest rates, and fees evolve while the credit agreement is active. Accordingly, we use five commercial databases and two linking databases to conduct our analysis. We also hand-collect data for a random sample of US publicly traded corporations that have at least one active credit line in the sample period. In this regard, first, we use 10-Q and 10-K US Securities and Exchange Commission (SEC) filings to obtain data on the quarterly usage of facilities at the credit line level. Although Standard & Poor's (S&P) Capital IQ provides data on usage, it does so at the firm level and does not always differentiate between revolving and term loans. Collecting data at the facility level, however, involves an obstacle: Large firms tend to have a considerable number of facilities simultaneously and usually disclose information about usage in an aggregate manner. To overcome this problem, we focus on firms with assets below $20 billion. This threshold makes sample firms comparable to mid- and small-cap firms. However, it does not seem a too restrictive a threshold, since less than 5% of non-financial US firms in S&P's Compustat universe have assets above $20 billion. Second, 10-Q and 10-K reports reveal that the main commercial database that provides data on lines of credit, Thomson Reuters Loan Pricing Corporation's DealScan, does not include a relatively large number of amendments to credit agreements (around 30% over the sample facilities). This flaw is extremely relevant to our analysis, which requires taking into account changes that could affect quarterly returns on facilities. Therefore, we use information in 10-Q and 10-K reports to find contracts missing from DealScan in the Electronic Data Gathering, Analysis and Retrieval (EDGAR) system. To obtain the entire set of sample contracts, we also collect the contracts of the facilities in DealScan. Third, we use this set of contracts to obtain information that DealScan does not cover or covers insufficiently.

Our primary result is that expected returns on lines of credit seem to carry a risk premium. The presence of such a risk premium is robust to alternative measures of risk and different specifications, including controlling for lenders' systematic risk. It is an outcome, however, that is at odds with the results of Asarnow and Marker (1995), who observe that return and risk are not positively related. Nevertheless, another of our results is in line with those of Asarnow and Marker (1995): We find evidence suggesting that the market for lines of credit underprices the risk of drawdowns increasing as borrower creditworthiness deteriorates. Analysis of the 2007–2009 financial crisis suggests that risk and return were negatively related during this period, indicating that debt contracts were not designed to adequately compensate for the risk increases during the crisis. This finding, which seems related to the sharp increase in corporate default probability in the downturn, is in line with works pointing out that the risk–return relation in equity markets might not be stable across the business cycle, vanishing or even reversing during recessions (Ghysels et al. 2014, 2016). Probit analysis also suggests that quarterly increases in risk augment the probability of observing increases in returns associated with credit line usage. This effect appears to be stronger if we focus on large increases in both risk and returns.

Although this paper builds on that of Asarnow and Marker (1995), their analysis has limitations that could help explain why they do not observe a positive relation between risk and return. We use information that they do not, such as facility-level data on usage and the determinants of the quarterly values of spreads and fees. Thus, our analysis is not performed at the level of a broad risk category but, rather, at the facility level and we do not assume constant levels of usage; that is, our data enable us to determine how returns evolve while facilities are outstanding and at the credit line level itself. Additionally, although we also perform a univariate analysis in the style of Asarnow and Marker (1995), we use a multivariate framework to examine the risk–return relation in the market for revolving facilities. To the best of our knowledge, this is the first paper to analyze this relation by means of multivariate econometric techniques. We expect, therefore, our work to help shed light on an almost unexplored area of research. Contributing to this aim and based on our hand-collected data, we also describe features of facilities' interest rate and fee schemes, to which academic works have paid little attention.

This paper relates to previous empirical studies examining the relation between corporate risk and contractual provisions on interest rates and fees in the market for credit lines (Shockley and Thakor 1997, Strahan 1999, Beatty et al. 2002, Asquith et al. 2005, Manso et al. 2010, Berg et al. 2016, Duran 2017). Instead of focusing primarily on those provisions, however, we focus on returns. In addition, in contrast to most previous work, we do not adopt an approach in which the relation between a contractual feature and risk is analyzed at origination: we examine the relation between the changing values of risk and returns while the facility is outstanding. Our research is also linked to the literature that uses option pricing theory to value credit lines (Bartter and Rendleman 1979, Thakor et al. 1981, Hawkins 1982, Thakor 1982, Ho and Saunders 1983, Chateau 1990). However, our analysis of the relation between risk and return does not require assumptions that, despite being inconsistent with commercial practices in the credit line market, are common in this literature.

Additionally, this research is closely related to studies that have analyzed how financial covenants protect lenders through restricting borrower behavior (Dichev and Skinner 2002, Chava and Roberts 2008, Nini et al. 2009, Sufi 2009, Demiroglu and James 2010, Demerjian and Owens 2016). Our paper complements these previous works in the sense of analyzing how risk and returns are traded below the risk threshold that covenants establish.

The remainder of the paper proceeds as follows. Section 2 constructs the econometric model and explains how we compute returns. Section 3 describes the data, sample construction, and summary statistics. Section 4 presents the empirical results. Section 5 examines the effects of the subprime crisis on the risk–return relation. Section 6 studies whether increases in risk modify the probability of expected returns also rising. Section 7 concludes the paper. Appendix A defines the variables used in the analysis. Appendix B gives details about the main features of credit lines' interest rate and fee schemes. Appendix C presents the criteria used to define the quarterly value of spreads and fees. Appendix D cites the purposes for which credit lines can be used according to credit contracts.

## 2. Model

A firm that enters into a revolving line agreement buys protection against uncertainty about whether it will face tighter lending conditions (Campbell 1978, Holmström and Tirole 1988). This insurance-like feature can intensify standard potential conflicts between borrowers and lenders. As long as no restrictions on access to the credit line, if any, are activated, the firm can still draw down as its quality deteriorates, even if it engages in risk-shifting strategies (Jensen and Meckling 1976, Chava and Roberts 2008). To reduce those conflicts and align the incentives of lenders and borrowers, credit contracts include provisions designed to protect the former. Specifically, interest rate and fee schemes are expected to be designed so that returns on credit lines compensate lenders for firm risk. In line with empirical findings indicating that lenders are risk averse, this risk-compensating role of returns would materialize in a trade-off between corporate risk and the expected return on the credit line. Thus, our hypothesis is that revolving lines of credit include a risk premium, consistent with risk-averse lenders trading expected returns for increased risk.

To examine whether empirical evidence supports this hypothesis, along with a univariate analysis à la Asarnow and Marker (1995), we estimate the following model:

$$E_t(Return_{i,t+1}) = \alpha + \beta \cdot Risk_{j,i,t} + \sum_{k=1} \gamma_k \cdot F_{k,j,i,t-1} + \sum_{h=1} \delta_h \cdot L_{h,i,t+1} + \theta \cdot Crisis + \varepsilon_{i,t}$$

In the base case, $E_t(Return_{i,t+1})$ stands for the return that is expected, in quarter $t$, to be yielded by facility $i$ in quarter $t+1$. The next section provides a detailed discussion of the concrete notion of return that we use and how it is computed. Following Santos (2011), Roberts (2015), and Berg et al. (2016) and in line with standard analysis of the risk–return relation in the financial literature, we capture corporate risk in the base model by the 12-month standard deviation of firms' daily stock returns. Specifically, $Risk_{j,i,t}$ is the standard deviation of the daily stock returns of firm $j$, holding line $i$, during the year that finishes at the end of quarter $t$. If facilities charge a risk premium that compensates for increases in borrower risk, that is, if our hypothesis is consistent with the empirical results, $\beta$ would be positive.

To test this prediction, our model controls for a set of firm- and credit line–specific variables: $F_{k,j,i,t-1}$ and $L_{h,i,t+1}$, respectively. In line, for instance, with Sufi (2007), Roberts and Sufi (2009), and Delis et al. 2016, firm characteristics are lagged by one quarter relative to the quarter in which the expectation of the return is formed. Facility characteristics are contemporaneous to the quarter in which the expected return would be yielded. To control for the effects of the subprime crisis, the model includes an indicator variable that equals one during the quarters of the crisis, which, according to the National Bureau of Economic Research (NBER), ranges from 2007:Q4 to 2009:Q2. Given that we use expected instead of current returns, this indicator variable is lagged by one quarter relative to the crisis period. Following Santos (2011) and Berg et al. (2016), we use dummy variables for firms' credit ratings, along with quarter (0, 1) indicator variables. To further control for firm-level heterogeneity, we include borrower fixed effects in the regressions. As part of the identification strategy, we also include lender fixed effects.[3] To control explicitly for potential risk diversification by the lender, we drop lender fixed effects from one of our robustness tests and include in the base model a variable that captures lenders' systematic risk. Finally, $\varepsilon_{i,t}$ is a random error term that, in line with Roberts (2015), is assumed to be correlated within facility observations and potentially heteroscedastic.

---

[3] Following, for example, Ivashina and Scharfstein (2010b), if a facility is syndicated, we control for the lead lender and identify it as the member of the syndicate designated as the administrative agent. Lenders are aggregated at the level of their top holding company.

Regarding firm-specific controls, we include six variables that capture the most frequent types of financial covenants found in dollar-denominated revolving lines of credit outstanding in the sample period and entered by non-financial US public firms; that is, we include variables that proxy for leverage, coverage, capital expenditures, net worth, current ratio, and profitability.[4] Controlling for how covenants can influence the risk–return relation is relevant because covenants can affect the usage of and, therefore, returns on lines of credit (Sufi 2009). As Appendix A shows, the way in which we define the variables capturing covenants is based on standard definitions of the covenants in credit contracts (Wight et al. 2007). The expected effect of these variables over returns is also based on facility contracts. Specifically, we expect a covenant variable to be positively (negatively) related to returns if, according to credit contracts, that covenant is breached when the variable violates an upper (lower) limit. Thus, we expect leverage, coverage, net worth, the current ratio, and profitability to be negatively related to returns,[5] whereas we expect a positive relation between capital expenditures and return.

We also include five more control variables for firm characteristics that can alter the risk–return relation: size, the market-to-book ratio, tangibility, financial constraints, and monitoring costs. Since larger firms are more diversified and less likely to file for bankruptcy, we expect size to have a negative effect on return. The market-to-book ratio serves as a proxy for firms' growth opportunities and, therefore, we expect an increasing market-to-book ratio to be positive for lenders and negatively related to return. Regarding tangibility, in the case of bankruptcy, firms with more tangible assets reduce lenders' losses; hence, we expect a negative relation between tangibility and return. Roberts and Sufi (2009) show that borrowers' options to obtain alternative sources of financing play an important role in the definition of credit contracts. To take into account these outside options, we use the linearization of Lamont et al. (2001) of the Kaplan–

---

[4] According to DealScan, 26% of dollar-denominated revolving facilities outstanding in the sample period and entered by non-financial US public firms include covenants based on the ratio of debt to earnings before interest, taxes, depreciation, and amortization (EBITDA) (38% include other versions of debt-related covenants), 35% include coverage covenants, 12% include capital expenditure covenants, 9% include net worth covenants, and 3% include current ratio covenants. Additionally, 70% include cash flow-based covenants. A similar distribution is obtained if we restrict the set of credit lines to sample facilities for which DealScan provides information on covenants. Chava and Roberts (2008) and Sufi (2009) provide also comparable data on the use of covenants among public corporations.
[5] As Appendix A shows, leverage is defined as the inverse of the debt-to-EBITDA ratio. We use the inverse of this ratio so the expected effect of this variable on return does not depend on whether EBITDA is positive or not. Specifically, we expect increases in leverage to reduce return.

Zingales (1997) index to control for the effect of firms' financial constraints. Given that lenders could demand higher returns from firms with higher monitoring costs (Strahan 1999), the empirical model also controls for these costs. In this regard, previous works have related reporting accruals to earnings inflation and the opacity of cash flows and, therefore, to the need for more rigorous monitoring (Teoh et al. 1988, Sufi 2007). Accordingly, following Sufi (2007), we capture monitoring costs by means of accruals, as measured by Sloan (1996).

With respect to credit line characteristics, we control for the commitment amount and the maturity stated in the credit contract and for whether lines can be used for corporate restructuring, are provided by a syndicate of lenders, and are secured. As Santos (2011) points out, facilities with larger commitments or longer maturities may face more credit risk but are more likely to be granted to high-quality firms. Hence, the relations between the commitment amount and maturity to return are ambiguous. The relation between the indicator variable for whether facilities are secured and return is also ambiguous: Although secured facilities are safer, lenders are more likely to require a facility to be secured if the borrower is not of sufficient quality (Berger and Udell 1990). We expect credit lines whose purpose is corporate restructuring to be used more heavily and, therefore, to generate higher returns. If a credit facility is syndicated, coordination costs rise and, hence, the return on the facility is also expected to be higher (Bolton and Scharfstein 1996, Asquith et al. 2005).

Additionally, we include (0, 1) indicator variables denoting for whether lines include commitment, annual, utilization, and upfront fees. A (0, 1) indicator variable is also used to control for financial covenant violations. Given that covenant breaches are damaging for lenders and, if no waiver is given, they imply higher borrowing costs (see Appendix B), we expect a positive relation between this dummy variable and return. Finally, since having or not having outstanding borrowings can make a substantial difference on returns and outstanding borrowings are zero for about half of the observations, we control for whether drawdowns are nonzero.

**2.1. Computing expected returns**

This paper tests the risk-compensating role of returns in the market for revolving facilities. Accordingly, we focus on returns that, based on data availability, allow us to capture how facility contracts define risk premiums. Specifically, the dependent variable of our base analysis is the expected annualized quarterly coupon return; that is, taking into account the probability of default and given the level of usage, it is the annualized return that interest rate spreads and fees yield quarterly, following the contractual terms and while the credit agreement is active.[6]

Given the complexity of credit line pricing and data availability, we make some assumptions to calculate returns. Although some of our assumptions follow those of Asarnow and Marker (1995), we have data on quarterly usage at the credit line level and detailed information on credit lines' interest rate spreads and fees. Thus, besides avoiding assumptions on usage and relaxing those that these authors made on pricing, our analysis is performed at the facility level and not on a broader category, such as major rating classifications; that is, in contrast to Asarnow and Marker (1995), we do not assume that the required spreads and fees of the (newly issued or not) facilities of a firm with a given level of risk are equal to the average spreads and fees of newly issued facilities of firms with the same risk. In addition, we do not assume that all firms with the same level of risk have an equal and constant level of facility usage.

We assume that loans are reset quarterly, because this is the frequency at which variable spreads and fees are actually modified and also because it is the frequency of our data on credit line usage. Indeed, we compute the value of quarterly returns at quarter-ends.

In all except for one fixed rate facility,[7] sample credit lines charge an interest rate on drawdowns equal to a base rate plus a spread. The former defines the types of loans that can be drawn down from credit lines. The standard differentiation that credit agreements make is between the London Interbank Offered Rate (LIBOR) and alternate base rate (ABR) loans. As Wight et al. (2007) and Duran (2017) point out, most credit lines give firms the chance to choose between borrowing either of these types and to convert one type to the other while loans are outstanding; specifically, 93.00% of sample facilities include this option.[8] Since we have no information on how usage distributes between ABR or LIBOR loans, we make an assumption about the chosen type and, hence, the applicable interest rate at each reset time. Following Asarnow and Marker (1995), we assume that borrowers are cost minimizers, that is, they borrow the type of loan with the lowest interest rate. If a credit line allows ABR and LIBOR loans but we just have information on the interest rate charged on one of these types, we use the latter rate. Similarly, if we cannot calculate the fee of a credit line in a quarter, we use available data on this facility's pricing. For those credit lines in which a lender's official prime rate is used to determine ABR, we assume that this rate equals the prime rate reported by *The Wall Street Journal*.

---

[6] Note that we do not take into account income generated by the base rates of interest rates, because base rates do not define risk premiums. In addition, due to lack of data availability, we do not consider price returns.

[7] This fixed rate line was entered into by Miller Energy Resources Inc. on June 29, 2012. In line with our data, the sample of Shockley and Thakor (1997) includes only 13 fixed rate agreements out of 2,526.

[8] This option can be illustrated with the credit agreement of Triquint Semiconductor Inc. dated September 30, 2010. The contract gives the borrower the option to select Eurodollar or base rate loans (in our wording, LIBOR or ABR loans) and to switch between them: "Each Borrowing, each conversion of Loans from one Type to the other, and each continuation of Eurodollar Rate Loans shall be made upon the Borrower's irrevocable notice to the Administrative Agent … Each Loan Notice … shall specify … the Type of Loans to be borrowed or to which existing Loans are to be converted."

With respect to fees, following Asarnow and Marker (1995), we do not take into consideration fees that are not distributed among all lender members of a syndicated credit line, such as syndication and agency fees, or fees that are marginally present (Berg et al. 2016). Specifically, we focus on the four main types of fees in revolving credit lines: commitment, annual, utilization, and upfront fees.[9]

To determine applicable spreads and fees when they are variable, we require detailed data on pricing criteria. Since DealScan's coverage in this regard has significant limitations, we obtain this information directly from credit contracts. Pricing criteria are shown in Appendix C. To maximize the level of accuracy in the computation of pricing criteria, we use available data from Compustat, DealScan, Capital IQ, and Datastream, as well as manually collected data. Our calculations are based on 51 pricing criteria. According to some credit agreements, applicable margins and/or fees are determined by more than one criterion. As Table C2 in Appendix C indicates, this is the case for 5.01% of sample lines with a pricing grid.

Upfront fees are single charges that borrowers pay at origination. As for the rest of the data on credit line pricing, we hand-collect data on upfront fees directly from credit agreements. Thus, for a significant amount of credit lines, we have obtained information on this type of fees that DealScan does not include.[10] Nevertheless, a number of credit agreements mention that upfront fees, if any, are accounted for in nonpublic fee letters. According to Berg et al. (2015), contacts within the credit industry give DealScan an advantage in access to this nonpublic information. Therefore, although a credit contract makes no reference to an upfront fee, we consider that the corresponding credit line includes such a fee if DealScan does.

---

[9] A cancellation fee is a one-time charge against early termination or commitment reduction. This type of fee is not activated in any of the sample credit lines that include it. Hence, early termination fees are irrelevant in determining returns on sample facilities.

[10] One of the reasons why DealScan insufficiently records the information on upfront fees available in credit agreements is that this database includes a relatively low fraction of amendment fees, that is, upfront fees paid by borrowers on amendment closing dates. In addition, although a large number of the credit lines included in DealScan's *facilityamendment* table have upfront fees, DealScan does not reflect this information.

We amortize upfront fees on a straight-line basis, that is, for a given upfront fee, the amount amortized each quarter is equal to the fee divided by the quarterly duration stated in the agreement. Since renegotiation through amendments is quite frequent, we consider that, for amortization purposes, an original contract is still in existence, with the maturity stated in the contract as the limit, through successive amendments. Thus, we assume that any upfront fee—included in either the original contract or any of the amendments in the loan path—can be amortized while an amendment is in effect and by the amendment in effect. Indeed, we take into account any upfront fee included in any amendment of the sample facilities, even if the amendment does not modify the principal, pricing, or maturity. We use alternative amortization schemes in robustness tests.

To calculate the income generated by commitment fees, we use the available amount to borrow under credit commitments. This amount is usually reported in 10-Q and 10-K filings but, if not, we compute it. In those cases in which we are dealing with a credit line that has a borrowing base (or a program to support the issuance of letters of credit) but we have no data on the borrowing base (or letters of credit outstanding) for a quarter, we assume that the borrowing base is equal to the total commitment (or there are no letters of credit outstanding).

Since the focus of our analysis is on how facility contracts define borrowers' risk premiums, the numerator of the return (conditional on no default) is the quarterly coupon income. The denominator equals outstanding borrowings plus the amount of capital that the lender must legally set aside to cover contingent exposure to the unused portion of the commitment. This amount of capital results from multiplying the available portion of the line of credit times a so-called credit conversion factor, whose value depends on the duration of the commitment. In the sample period, this factor is 0.5 if the maturity of the facility is longer than one year and zero otherwise.[11] However, since the Code of Federal Regulations establishes 14 months as the maturity threshold that makes the conversion factor equal to zero, we perform a robustness analysis where the latter threshold is used to compute the denominator of return. In addition, given that the denominator of the return is zero for facilities with no outstanding borrowings and a maturity below one year (or 14 months in the robustness check), we also test whether our results are robust to a conversion factor of 0.5 for all facilities.

---

[11] See Section D, Conversion Factors for Off-Balance Sheet Items, of the Federal Deposit Insurance Corporation Rules and Regulations, Appendix A to Part 325, Statement of Policy on Risk-Based Capital.

To factor expected default losses into returns and, thus, obtain expected coupon returns, we use the firm's probability of default, measured as Bharath and Shumway (2008) propose, based on Merton's (1974) distance-to-default model (see Appendix A for further details). To compute the expected return and due to data unavailability, we assume that, in the state of nature in which a firm defaults, there is a standard loss-given-default rate. Specifically, following Asarnow and Marker (1995) and in line with previous estimations of this rate, we assume that it is equal to 34.8% over the promised coupon return.[12]

The literature on credit lines standardly distinguishes between all-in spread drawn (AISD) and all-in spread undrawn (AISU) spreads and fees (Strahan 1999, Santos 2011, Berg et al. 2016, Duran 2017). We also use this distinction to enrich the analysis of the risk–return relation. Specifically, we decompose the total expected coupon return into two components. The first, the expected AISD return, is the yield associated with credit line usage, that is, the return generated by the applicable spread of interest rates and by annual and utilization fees. The expected AISU return is associated with the right to draw down on credit lines. It is the result of the income generated by annual, commitment, and upfront fees.[13]

**3. Data and sample construction**

This section describes the sample construction process, as well as the basic characteristics of the sample data. We also compare our sample to the Compustat and DealScan databases.

**3.1. Sample construction: The starting dataset**

The final dataset of our analysis is the result of merging information from five commercial databases and a manually collected dataset. We extract data on credit facilities from DealScan and quarterly accounting data on firms from Compustat. As Appendix C indicates, data from Capital IQ and Thomson Reuters Datastream are used in the computation of some of the criteria determining applicable interest rate spreads and fees. Data provided by the Center for Research in Security Prices US Stock Database (CRSP) are used to calculate the standard deviation of sample firms' stock returns. We merge this database and Compustat by means of the CRSP/Compustat Merged Database.

---

[12] In a study of 24 years of defaulted commercial and industrial loans at Citibank, Asarnow and Edwards (1995) find the loss-given-default rate to be equal to 34.8%. Using Moody's data, Gupton et al. (2000) observe that the mean bank loan value in default is 69.5% for senior secured loans and 52.1 for senior unsecured loans. Since 63.29% of our sample lines are secured, the loss-given-default rate applicable to our sample according to these authors is 36.89%, very close to the markdown that we use. Based also on Moody's data and holding seniority constant at the senior secured level, Schuermann (2004) reports that the loss-given-default rate for bank loans is 36.90. Analyzing 18 years of loan loss history at JP Morgan Chase, Araten et al. (2004) conclude that the mean value of this rate is 39.8%.

[13] Although these two concepts of return are closely related to the return that would be yielded by DealScan's AISD and AISU, there are differences between them. Regarding AISD, on the one hand, DealScan does not take into account utilization fees, and, on the other hand, it assumes that the spread is that of the LIBOR and not the spread of the interest rate applicable under the assumption that borrowers are cost minimizers. With respect to the AISU, DealScan does not include upfront fees.

Our first step is to merge the DealScan and Compustat data. To do so, we use the linking database provided by Chava and Roberts (2008). This step allows us to obtain a dataset with the quarterly accounting data of firms that have at least one outstanding revolving credit line in the sample period, 2006:Q1 to 2012:Q2. Quarterly data are used due to the high frequency of credit agreement renegotiations, which makes changes in pricing schedules quite common in the life of credit facilities (Roberts and Sufi 2009, Godlewski 2014, 2019, Roberts 2015, Nikolaev 2016). The sample period ends in 2012:Q2, because this is the last quarter covered by the DealScan–Compustat linking database. The main reason for using 2006:Q1 as the starting quarter of our analysis is the time cost involved in manually collecting information on the usage and features of credit lines. Our sample, nevertheless, extends through 26 quarters.

Once the starting DealScan–Compustat dataset is generated, we apply filters. We exclude firms in DealScan that cannot be matched to those in Compustat, as well as non-US and financial firms.[14] In addition, we only keep observations corresponding to dollar-denominated revolving lines of credit.[15] Following Sufi (2009), we also require firms in the DealScan–Compustat dataset to have a minimum number of consecutive quarters in Compustat, with active lines of credit.[16] This condition is established to reduce the probability of credit lines with no observations in the final, randomly chosen dataset.

---

[14] To exclude financial firms, we drop firms with SIC codes 6000–6999.
[15] Following Berg et al. (2015), we select loan commitments whose DealScan variable *loantype* is either *Revolver/Line < 1 Yr.*, *Revolver/Line >= 1 Yr.*, *364-Day Facility*, *Limited Line*, or *Revolver/Term Loan*.
[16] Sufi's (2009) condition is more restrictive: It requires firms to have at least four consecutive years (in an eight-year period) of positive data for the main variables of the analysis. To describe our condition in a more detailed manner, let us define the concepts *starting quarter* and *ending quarter* by means of an example. Consider a firm that is included in Compustat between 2006:Q2 and 2010:Q2 and that has two lines of credit active in the sample period 2006:Q1 to 2012:Q2. The origination (maturity) quarters of these lines are 2006:Q3 (2007:Q3) and 2007:Q2 (2103:Q2), respectively. In this case, the starting quarter is the latest between 2006:Q2 or 2006:Q3, that is, the latest quarter between the first quarter in which the firm is included in Compustat (2006:Q2) and the earliest quarter among the origination quarters of the facilities the firm has in the sample period (2006:Q3 between 2006:Q3 and 2007:Q2). If the latest quarter between the candidates 2006:Q2 and 2006:Q3 is earlier than 2006:Q1, the starting quarter would be 2006:Q1. The ending quarter is the earliest between 2010:Q2 and 2013:Q2, that is, the earliest quarter between the last quarter in which the firm is included in Compustat (2010:Q2) and the latest quarter between the maturity quarters of the facilities the firm has in the sample period (2013:Q2 between 2007:Q3 and 2013:Q2). If the earliest quarter between the candidates 2010:Q2 and 2013:Q2 is later than 2012:Q2, the end quarter would be 2012:Q2. Given these definitions, we require firms to have at least four consecutive quarters between the starting and ending quarters.

Large corporations tend to have a relatively high number of lines of credit active simultaneously and usually report on their usage in an aggregate manner. This aggregate information does not allow for computing returns at the facility level and, hence, is inadequate for our research goal. Accordingly, to make it possible to hand-collect disaggregated data, we exclude from the DealScan–Compustat dataset firms with an asset book value above $20 billion in any sample quarter. This $20 billion threshold is based on the fact that the maximum asset value of a firm included in the S&P MidCap 400 or SmallCap 600 indexes during the sample period amounts to $19.921 billion. Therefore, since sample firms are not necessarily part of these stock indexes, any bias associated with being listed is avoided, but the companies listed in these indexes resemble sample firms in terms of maximum size. The threshold, nevertheless, does not seem too restrictive: Among non-financial US companies included in Compustat in the sample period, 96.37% have an asset size below $20 billion during the whole period.

The DealScan–Compustat dataset has 206,883 facility–quarter observations that refer to 8,908 lines of credit and 2,545 firms. We randomly sample 150 firms from this dataset. Our manual data collection process is based on these firms' lines of credit. In line with Sufi (2009) and Roberts (2015), we restrict the number of randomly selected firms because manually collecting the required data implies reviewing a large number of credit contracts and 10-K and 10-Q reports. Even for 150 firms and 26 quarters, this process implies looking over several thousand SEC filings.

**3.2. Sample construction: The manually collected dataset**

The aim of manually collecting data is to obtain information on aspects of lines of credit that are not adequately covered by commercial databases. Specifically, we address three main gaps. Regarding the first one, SEC regulation compels firms to provide detailed information about their credit lines in 10-Q and 10-K reports (Kaplan and Zingales 1997, Sufi 2009). These reports reveal a relevant limitation of DealScan: Firms refer to amendments to existing lines of credit—and occasionally to newly originated facilities—that this database does not include. Therefore, the information provided by this commercial database could be insufficiently accurate for analyzing the risk–return relation: Amendments to the pricing schedule or to any other relevant aspect of a facility could have been agreed upon without DealScan reflecting these changes. To overcome this limitation, we search for amendments—or new revolving lines of credit—not covered by DealScan in the list of exhibits that appears at the end of 10-Q and 10-K reports. Reference to an exhibit in this list is usually complemented by information that allows us to locate the original credit contract in EDGAR.[17] Once we find the agreement, we include it in our dataset if, as Roberts and Sufi (2009) require, it refers to a new line of credit or amendment that does not leave unchanged the principal, interest rates on drawdowns, fees, or maturity.[18] Any information referring to manually added facilities is hand-collected directly from the credit contracts themselves. We also collect the credit contracts of sample facilities covered by DealScan. The latter contracts are used to obtain information about features of facilities not covered or insufficiently covered by DealScan.

Manually added facilities constitute almost a third (30.86%) of all the lines of credit in our dataset. The vast majority (90.67%) of these added lines were amendments and amended and restated agreements.[19]

The second gap that we cover by manually gathering data refers to information not provided by commercial databases. In this respect, we collect data on credit line usage and availability at the facility level from 10-K and 10-Q reports. The only commercial database that provides this type of information is Capital IQ, but it does so at the firm level and does not always differentiate between revolving lines of credit and term loans.[20]

An example of how firms disclose their credit line usage is provided by Arch Coal Inc. in its 2011 10-K filing: "As of December 31, 2011, we had borrowings of $375 million under our $2 billion dollar revolving credit facility." Nevertheless, although public firms must legally report on their credit lines, there is no explicit requirement for disclosing facility usage. Therefore, we drop any quarter–facility observation for which we find no data on outstanding borrowings. Similarly, we drop facility–quarter observations for which, despite the $20 billion threshold on corporate asset size, a firm reports outstanding borrowings under its facilities in an aggregate manner.

---

[17] For instance, the 2011 10-K filing of Moog Inc. lists the following exhibit: "Third Amended and Restated Loan Agreement … dated as of March 18, 2011, incorporated by reference to exhibit 10.1 of our report on Form 8-K dated March 18, 2011." Accordingly, the credit agreement can be found as exhibit 10.1 and attached to an 8-K report dated March 18, 2011. This exhibit was filed a few days later, on March 21, 2011. Sometimes, however, the search process is not so straightforward, mainly because the list of exhibits does not provide information about the filing date and this date is not close to the date of the report to which the agreement is attached. In these cases, a large number of filings around the origination date must be consulted before the credit contract is found.

[18] Our method, however, is more exhaustive than that of Roberts and Sufi (2009). Their conclusion about whether a renegotiation has taken place depends on whether the firm discloses changes in the features of the credit agreement in 10-Q and 10-K reports, whereas we search for these changes directly in the credit contracts.

[19] Our dataset also includes what we call *artificial facilities* (8.87%). We introduce this category to reflect situations where the principal of a credit line changes not through an amendment but according to provisions included in the credit agreement itself. For instance, the fifth amendment to a loan agreement dated December 21, 2007, and entered into by Cascade Corporation includes the following clause: "The Aggregate Commitments shall be reduced by $1,250,000 on a quarterly basis beginning on March 31, 2008, and continuing on the last day of each subsequent quarter for so long as this credit facility is active."

[20] For instance, according to the variable *activebalrrevolvingcredit* of Capital IQ, Time Warner Inc. had $12.381 billion of revolving credit active at the end of 2006. Nevertheless, the 10-K filing indicates that this amount corresponds to four different loan commitments; that is, data are aggregated at the firm level. Moreover, two of those loan commitments were terms loans. For more details on the features of Capital IQ, see Manakyan and Giacomini (2016).

We also manually gather information on other features of facilities that are not covered by commercial databases. If a facility has a borrowing base, we collect data on its quarterly value if available. We generate a variable that indicates the date at which a credit contract stops being active either because it matures or for any other reason, such as its amendment or cancellation. Thus, we can compute the actual duration of any contract and can rebuild the entire loan path of any original contract, that is, any chain of amendments following the origination of a revolving facility and ending in a terminal event (Roberts 2015). Additionally, we collect data on whether a firm is in technical default on a facility in a given quarter as a result of violating a financial covenant. To compute returns on facilities as accurately as possible, we also check 10-K and 10-Q reports to determine whether firms are granted waivers that remove increases in fees or rates applicable under technical default.

To address the third gap, we hand-collect data that are insufficiently covered by commercial databases. As Roberts and Sufi (2009) point out, DealScan's coverage of data on pricing schedules has limitations. Accordingly, we collect data that allow us to determine the quarterly values of fees and interest rates; specifically, we collect data on the types of base rates, spreads and fees, any criterion or margin involved in determining applicable base rates, the values of the spreads and fees if fixed, the pricing criteria and pricing grids if variable, and other provisions that could modify applicable interest rates or fees such as those referring to technical default. On the basis of these hand-collected data, we discuss the main features of facility pricing in Section 3 and Appendix B.

Regarding the purposes of facilities, DealScan includes the variables *primarypurpose* and *secondarypurpose*, which refer to what seem to be the main and auxiliary purposes, respectively, for which credit facilities can be used. Nevertheless, typical credit agreements, first, do not establish any priority among purposes and, second, frequently include more than two purposes.[21]

---

[21] For instance, the credit agreement of Pacific Sunwear of California Inc., dated September 14, 2005, states, "The proceeds of the Loans will be used only to refinance certain existing credit facilities of the Borrower and to finance the working capital needs, capital expenditures, acquisitions (including Permitted Acquisitions), dividends, distributions and stock repurchases, and for general corporate purposes of the Borrower and its Subsidiaries."

Accordingly, we hand-collect data on credit line purposes. This information is shown in Appendix D. Around half of the sample credit lines (50.48%) have more than two purposes and only 17.70% have a single purpose. To deal with the complexity resulting from this variety of purposes, we follow Ivashina and Scharfstein (2010a): We split credit lines between those that could be used for corporate restructuring—leveraged buyouts, mergers and acquisitions, and stock repurchases—and the remainder.

For the sample facilities included in DealScan, we check the data that this database provides on principal and maturity with credit contracts. Besides reducing the probability of potential errors, this comparison avoids situations where, for instance, credit line usage is greater than the facility principal. The percentages of credit contracts that differ from DealScan in terms of commitment amount and maturity are very similar (6.16% and 6.51% of the sample facilities in DealScan, respectively).[22] Additionally, when the secured/unsecured status of a facility is missing from DealScan, we check whether contracts provide this information. This is the case for just 2.05% of the sample facilities in DealScan.

The last step in the construction of the dataset is to drop facility–quarter observations for which available information does not allow us to compute income generated by applicable fees and interest rate spreads. This condition yields our final sample: an unbalanced panel of 2,073 facility–quarter observations that includes 486 facilities.

**3.3. Summary statistics**

Table 1 presents summary statistics for facility features other than pricing and return. All the variables are defined in Appendix A. The statistics are calculated as if there were one observation per facility, except for technical default and zero outstanding borrowings, whose statistics are computed over firms and the total number of observations, respectively. Approximately two-thirds of the sample facilities (67%) are secured and most (90%) are syndicated. The average sample line has a stated maturity slightly longer than 44 months and a principal of approximately $266 million. Corporate restructuring is among the purposes of half of the sample facilities, outstanding

---

[22] Four sample facilities (with DealScan unique identifiers 175829, 184660, 208598, and 226103) modify their principal either between origination and the end of the origination quarter or before 2006:Q1. We directly use the modified principal as the commitment principal but do not consider these cases among those in which DealScan and credit contracts differ.

borrowings are not zero for approximately half (52%) of facility–quarter observations, and the rate of firms violating covenants in our dataset is 17%.

**[Table 1]**

For comparison purposes, Table 1 also presents statistics for revolving lines of credit in DealScan, denominated in US dollars, active in the sample period and made to non-financial US firms. These facilities are quite similar to those in our sample, although the percentages of secured and syndicated lines are lower in the sample lines (12% and 8% less, respectively). In addition, sample facilities have a slightly shorter average maturity (six months shorter) and almost the same average commitment.

**[Table 2]**

Table 2 presents summary statistics of the main characteristics of sample firms and non-financial US firms in Compustat. Relative to the latter database, our sample contains firms that are, on average, less highly levered, more profitable, and, according to the Kaplan–Zingales (1997) index, less financially constrained. These firms also have slightly greater asset tangibility, a better coverage ratio, and lower monitoring costs. However, the sample firms have lower current and market-to-book ratios, have lower net worth, and are smaller. The capital expenditures of the sample and Compustat firms are about the same. Overall, these differences are consistent with our sample selection criteria, since, on the one hand, we require firms to have an asset size below $20 billion in all sample quarters and, on the other hand, we focus on firms that have at least one outstanding credit line in the sample period. Indeed, the differences that we find between Compustat and our sample firms are in line with those observed by previous papers whose empirical exercise is based on the intersection of Compustat and DealScan data or a random subsample taken from this intersection (Chava and Roberts 2008, Roberts and Sufi 2009, Roberts 2015).

With respect to interest rates and fees, our dataset allows this paper to be the first to not only take into account their values at origination, but also identify how they evolve while credit lines are outstanding. Appendixes B and C provide details regarding the interest rate and fee structure of revolving facilities.

**[Figure 1]**

Figure 1 shows how the AISD and AISU of the sample lines, whose averages are 179.79 bps and 31.30 bps, respectively, change during the sample period. The plot indicates that the costs associated with usage and the right to use facilities follow very similar trends through the sample period. Just before the 2007 crisis begins, in 2007:Q4, the AISD and AISU start an almost uninterrupted increase that continues after the end of the crisis and peaks in 2011:Q2. In this quarter, the AISD and AISU equal 246.67 and 39.92, respectively; that is, they have increased by 85.77% and 54.97%, respectively, from their troughs in 2007:Q2.

## 4. Results

In this section, we present the empirical results of our analysis of the risk–return trade-off in the market for revolving credit lines. First, in line with the analysis of Asarnow and Marker (1995), we show the outcomes of a univariate study. Second, we discuss the findings obtained from our multivariate econometric approach. Finally, to check for the robustness of our results, we perform a series of additional tests.

### 4.1. Risk premium: Univariate analysis

Table 3 synthesizes the results of the univariate analysis of the relation between credit facilities' returns and firm risk. It shows the average expected annual coupon return on revolvers for the whole sample and by risk category.[23] As Santos (2011), Roberts 2015, or Berg et al. (2016), we use the standard deviation of firms' daily stock returns to proxy for corporate risk. Accordingly, risk categories are given by the quintiles of this standard deviation. We have also computed the mean expected annual AISD and AISU returns.

In their analysis, Asarnow and Marker (1995) find evidence of mispricing in the facility market. Specifically, Exhibits 4 and 7 of Asarnow and Marker (1995) suggest no trade-off between risk and return. Indeed, higher-risk firms yield lower mean annual returns on revolving facilities, whereas lower-risk firms generate higher returns. However, as the first row of Table 3 shows, our univariate analysis suggests that risk and return are positively related: From a value of 0.79% for the credit lines of firms in the lowest risk category, the expected return increases category by category, up to the highest, where the total return equals 1.67%.

**[Table 3]**

The second row of Table 3 indicates that a similar pattern characterizes the expected AISD return. Therefore, there also seems to be a positive relation between risk and the return associated with facility usage. However, as the third row of the table suggests, there is some ambiguity in the relation between risk and the expected AISU return, that is, the expected return yielded by the right to use facilities. Although, within a narrow range of variation, this return steadily increases across the first four risk categories, from 0.41% to 0.65%, it then decreases to 0.55%. Such ambiguity regarding the relation between risk and the expected AISU return raises concern about whether credit lines underprice the borrower's option to draw additional funds under a liquidity shock, as Asarnow and Marker (1995) have already pointed out. This conclusion will be further analyzed in the following section.

---
[23] To obtain the mean expected annual return for either the whole sample or each risk category, we calculate the average expected return of the sample facilities per quarter. Then, we compute the compound cumulative return. The mean annual return is the geometric average of this cumulative return.

**4.2. Risk premium: Regression analysis**

As with other debt instruments, revolving lines of credit are a source of risk for the lender. In line with empirical evidence indicating that the latter is risk averse (Ratti 1980, Hughes et al. 1995, 1996, 1999, Hughes and Moon 1997, Angelini 2000), we can expect revolving facilities to carry a risk premium. Other provisions included in the credit agreement, such as covenants, restrict the maximum level of risk that borrowers can take. Indeed, financial covenants protect the lender in a binary way, that is, if the borrower's performance triggers a covenant and no waiver is granted, then control rights revert to the lender (Aghion and Bolton 1992, Chava and Roberts 2008). However, risk premiums make it possible for lenders to accept higher levels of risk before covenant thresholds are reached. Moreover, since applicable rates and fees usually increase under technical default, returns can still play a compensating role in this situation (Asquith et al. 2005).

If returns play such a risk-compensating role, we expect return and risk to be positively related. Our investigation into this hypothesis is based on the model presented in Section 2. In line with previous research (e.g., Sufi 2007, Roberts 2015), to take into account the assumption that the error term of our regression equation is correlated within facility observations and potentially heteroscedastic, we estimate the effect of corporate risk on revolving facilities' returns using ordinary least squares regressions clustered at the facility level. We use four different econometric specifications, and the estimation results are displayed in Table 4. To examine the risk-compensating role of expected returns without controlling for any variable potentially affecting this role, the first specification (column (1) in Table 4) includes only fixed effects for the borrower, firm credit rating, quarter, and lead lender. The second specification (column (2)) corresponds to the base model discussed in Section 2. This model, along with the fixed effects taken into account in the first specification, includes firm- and credit line–specific control variables and an indicator variable for the 2007:Q4–2009:Q2 crisis period. The last two empirical specifications (columns (3) and (4)) are similar to the base model, but their dependent variables are the expected AISD and AISU returns, respectively. The former is associated with credit line usage; that is, this portion of the coupon return compensates the lender for the risk associated with the borrower's takedowns. The expected AISU return results from the amount the borrower pays for each dollar available under a commitment; that is, it captures the return for the facility being usable. Thus, it is expected to compensate for the risk arising from the borrower increasing usage as the borrower's creditworthiness deteriorates (Berg et al. 2016, Duran 2017).

According to columns (1) and (2) in Table 4, the credit facilities of riskier borrowers appear to yield higher expected coupon returns, whether or not the analysis incorporates control variables that could affect the risk–return trade-off; that is, our results suggest that returns include a risk premium that compensates lenders for increases in their borrowers' risk. Regarding the size of this compensation, the coefficient estimate of risk in column (2) (7.648) suggests that a 1% increase in the standard deviation of corporate daily stock returns causes a 0.254% increase in the return on credit lines at the mean, from 1.045% to 1.048%.[24]

**[Table 4]**

Column (3) in Table 4 indicates that there also seems to be a positive relation between expected AISD return and risk. Such an outcome suggests that the expected return generated by outstanding borrowings on the credit line compensates the lender for increases in the borrower risk. Indeed, the effect of risk over the return component associated with facility usage appears to be greater than for the whole return (9.857 vs. 7.648). In relation to the expected AISU return, column (4) shows no statistically significant relation between this component of the return and corporate risk. This result indicates that the market for lines of credit could be underpricing the risk of increased usage as corporate credit quality worsens. In this regard, Figure 2 illustrates the potential negative relation between corporate quality and facility usage. According to this figure, the average usage-to-commitment ratio increases across the quartiles of the distribution of corporate risk, with an especially intensive rise between the third and fourth quartiles, from 22.03% to 33.04%. The finding that riskier borrowers' right to draw down from revolving facilities could be underpriced, already suggested by our univariate analysis, is indeed in line with Asarnow and Marker (1995).

---

[24] The study of Asarnow and Marker (1995) does not provide quantitative results comparable to ours and there is no other analysis on the relation between risk and return in the market for credit lines. However, our outcomes can be related to research analyzing the relation between corporate risk and DealScan's AISD, as long as three caveats are taken into account: (i) Interest rate spreads and fees affect but are not equivalent to returns, (ii) DealScan's AISD does not include fees (utilization, upfront, or commitment) whose income we factored into return, and (iii) previous research uses data on pricing at contract origination, disregarding how spreads and fees change while contracts are outstanding. Bearing in mind these caveats, a trade-off in which a 1% rise in the standard deviation of corporate daily stock returns increases returns by 0.254% is in line with previous work. In this respect, according, for instance, to the results of Cai et al. (2012), the same rise in risk increases DealScan's AISD by 5.034 bps.

[Figure 2]

Overall, there do not appear to be significant inconsistencies among our priors and the estimated coefficients of the control variables. Regarding firm-specific controls, column (2) of Table 4 suggests a number of stylized facts. First, as expected, the returns are decreasing in the current ratio, profitability, and tangibility. The negative effect of these firm characteristics on returns is also found when the dependent variable of the analysis is the AISD return. However, the AISU return increases in tangibility. Second, capital expenditures increase total coupon and AISD returns. Third, given that leverage is captured by the inverse of the ratio of total debt to operating income, the AISU return is decreasing in leverage. It is also negatively related to size but increases in monitoring costs.

In relation to facility controls, maturity and returns are negatively related. A similar result is observed for AISD returns. Following Santos (2011), we argue these outcomes suggest that longer maturities are granted to more creditworthy firms. Syndicated facilities seem to compensate for their higher coordination costs by yielding higher returns (Bolton and Scharfstein 1996, Asquith et al. 2005). In addition, being a secured facility has a positive effect on returns. This result is consistent with lenders requiring riskier borrowers' facilities to be secured. Syndicated and secured facilities appear to also yield higher AISD and AISU returns, respectively.

Regarding fees, returns rise if the pricing scheme includes annual and commitment fees, whereas they seem unaffected by the inclusion of utilization and upfront fees. Although commitment fees are not used to define the AISD return, they have a positive effect on it. This positive relation is consistent with multiple-fee pricing schemes being a self-selecting device of borrowers. As Thakor and Udell (1987) show, a borrower with a high takedown probability is willing to self-select into a contract with a commitment fee, which are charged on the unused portion of the commitment. In this sense, the usage-to-commitment ratio is higher for sample facilities with a commitment fee (25.25% vs. 17.43%). This higher usage leads to a higher AISD return and helps explain, therefore, the positive relation between this return and facilities including a commitment fee. In contrast with the effect of commitment fees on the AISD return and despite the screening role of pricing schemes, we do not observe a negative relation between commitment fees and the AISU return. The reason is that these fees are the main source of such returns. A utilization fee is charged on the drawn amounts if usage exceeds a pre-established threshold. Therefore, this type of fee could have a negative effect on facility usage and, hence, help explain why facilities including such fees yield lower AISD returns. Indeed, the average usage-to-commitment ratio in facilities without utilization fees more than doubles this ratio in the remainder of the credit lines (23.74% vs. 10.89%).

Having nonzero outstanding borrowings increases returns but has different effects on the AISD and AISU returns: in tune with their raison d'être, using a facility increases the former and decreases the latter. Finally, returns decrease during the subprime crisis. In Section 6, we analyze the effect of the crisis period on the risk–return relation in more detail.

**4.3. Robustness checks**

To check for the robustness of our results regarding the risk premiums of revolving facilities, we perform a series of additional analyses. The main estimated coefficients of these tests are displayed in Table 5.[25]

**[Table 5]**

In our base approach, corporate risk is proxied by the volatility of firms' daily stock returns. As Santos (2011) or Roberts (2015) point out, this volatility is a market-based, forward-looking measure of the firm's risk. These two features make it particularly suitable to examine the relation between expected returns on credit lines and corporate risk. Nevertheless, to check whether our results are robust to a different measure for risk, we repeat our analysis using an alternative, book-based proxy; specifically, following Sufi (2009), we use Altman's (1968) Z-score to capture the level of corporate financial distress. Since the Z-score is an inverse measure of financial risk, its relation to return is expected to be negative.

The base model includes lender fixed effects. Given this identification strategy, we do not take into account concrete lender characteristics such as whether lenders follow strategies to diversify risk. To explicitly control for this aspect of lenders' risk management, the second robustness check drops lender fixed effects from the analysis and includes lenders' systematic risk.

---

[25] Complete results are available upon request.

Such inclusion implies losing around 16% of our observations. We capture systematic risk by means of lenders' beta, which is computed by regressing corporate excess returns on market excess returns.

To factorize income from upfront fees into coupon returns, we amortize the former on a straight-line basis. Nevertheless, there are other options to amortize this fee. The results in rows (3) and (4) of Table 5 are based on two alternative methods. In row (3), the quarterly amortized amount is equal to the upfront fee divided by the number of quarters between the quarter in which the contract (either the original contract or the amendment including the fee) is settled and the earliest quarter between the quarter corresponding to the maturity date of the contract and the quarter in which the loan path of the facility terminates. In row (4), upfront fees are amortized just while the credit contracts establishing them are not amended or terminated. To compute the returns upon which the results in row (5) are based, we use 14 months as the maturity threshold that divides facilities into those that do not have to hold capital for the unused portion of the commitment and those that do. In row (6), no maturity threshold is used to compute the returns.

Following Roberts and Sufi (2009), we compute firm-specific control variables as the four-quarter rolling average of each control. Nevertheless, the literature on lines of credit has proposed an alternative way to smooth potential seasonal patterns (Demerjian and Owen 2016). The estimated coefficients in row (7) of Table 5 result from his alternative method. In particular, we calculate corporate control variables using current stock and annualized flow variables. To annualize the latter, we sum the values of the flow variables in the current and prior three quarters.

The size of the effect of risk on return could be different at different levels of risk. If this were the case, our linear approach to the analysis of the risk–return trade-off would be inaccurate. To explore these potential nonlinearities in the relation between risk and return, in row (8) of Table 5, we add a risk-squared term to the base model.

Despite their widespread use, Compustat data raise concerns over the possibility of results biased by this database dropping failed, acquired, or delisted firms and, hence, keeping only survivor firms (e.g., Boyd et al. 1993). Although we cannot completely rule out that the analysis is not biased by this survivor effect, we check whether we can expect this potential bias to not alter our results significantly. With this aim, we test whether the positive relation between risk and return is also observed for the riskiest sample firms, which are more likely to be dropped from Compustat and, therefore, to give rise to the survivor bias. We define this set of firms as those whose mean standard deviation of stock returns is above the median of all sample firms. In rows (9) and (10) of Table 5, these means and medians are calculated taking into account the whole sample and the 2007 crisis periods, respectively.

Overall, the results from the base analysis are not qualitatively different from the robustness test results displayed in Table 5. Specifically, all robustness checks are consistent with the conclusion that corporate risk is positively related to expected total and AISD returns but has no significant effect on expected AISU returns. Additionally, the risk-squared term in row (10) is nonsignificant. Hence, the size of the change in return as risk is modified does not appear to depend on the level of risk. According to rows (9) and (10), a positive risk–return relation appears to also be observed for the set of high-risk firms. Therefore, despite the use of Compustat data raising relevant concerns about survivor bias, we do not expect this potential problem to significantly affect our results.

**5. The subprime crisis**

The risk–return relation in stock markets may be time variant (Lettau and Ludvigson 2010, Aslanidis et al. 2016); specifically, there may be no relation or even a negative relation during financial downturns such as the subprime crisis (Ghysels et al. 2014). To analyze whether this kind of outcome is also present in the market for credit lines, we examine whether this crisis modifies how risk and return relate to each other in this market. Indeed, the 2007–2009 recession was an extraordinary time for financial intermediation that considerably altered the corporate loan market (Ivashina and Scharfstein 2010a).

To control for the potential effects of the recession over the risk–return relation, we already include an indicator variable for the crisis in the base model. We now incorporate an interaction term in the base analysis to obtain evidence of whether the subprime crisis affected the risk-compensating role of returns. Panel A of Table 6 shows the key results obtained from this empirical exercise.[26]

**[Table 6]**

The qualitative features of the estimated coefficients of risk do not change when the interaction term between risk and the crisis is included in the analysis. In this sense, the coefficient of risk is significant and positive when the dependent variable is the total expected coupon return. Hence, a risk premium appears to be charged on facilities outside of the crisis period, that is, when the indicator variable standing for the crisis equals zero. Indeed, according to the coefficient of risk in column (1) of Panel A, outside the crisis a 1% increase in the sample standard deviation of risk increases the coupon return at the sample mean by 0.376%, which is greater than the rise observed when no interaction term is included in the model (0.254%). In regard to the other specifications, the coefficient of risk when the dependent variable is the expected AISD return is also significant and positive (column (2) of Panel A), and the coefficient is nonsignificant if the dependent variable is the expected AISU return (column (3) of Panel A).

---

[26] Complete results are available upon request.

The coefficients of the crisis are nonsignificant in all the columns of Panel A, but the coefficients of the interaction terms are negative across the specifications. Moreover, the coefficient of the interaction term in the specification corresponding to total expected returns is larger in absolute value than the coefficient of risk (-14.440 vs. 11.330). Thus, the effect that an increase in risk causes on the expected coupon return seems not only smaller during the subprime crisis, but also negative; that is, in line with previous works on equity markets, the risk–return relation appears to have reversed in the market for corporate facilities during the downturn, with returns decreasing in risk.

Along with the type of mispricing discussed in previous sections (that associated with returns not satisfactorily compensating for the risk of distressed firms increasing facility usage), these findings suggest a second type of mispricing. Specifically, credit contracts appear to be inadequately designed to yield a risk premium that compensates for the increases in risk that took place during the crisis. However, we do not observe a similar reversion of the trade-off between risk and returns if the latter is measured by the expected AISD return. Although the coefficient of the interaction term between risk and the crisis is negative, the absolute value of this coefficient is smaller than that of risk; that is, risk is positively related to expected AISD returns in the downturn, although the effect of the former on the latter is less intense than in the remainder of the sample period.

Regarding expected AISU returns, the significance and negative sign of the coefficient of the interaction term in column (4) of Panel A, along with the nonsignificance of the coefficient of risk, indicates that risk affects this return differently within and outside of the crisis period. Risk and AISU returns do not seem to be significantly related outside of this period, but the former appears to decrease the latter during the recession. Thus, including an interaction term in the base model suggests that the two types of mispricing found in our analysis are present in the relation between risk and expected AISU returns. On the one hand, this return decreases in risk during the 2007–2009 crisis. On the other, outside of the downturn, AISU returns do not adequately compensate for the risk of troubled firms use their credit lines more intensively.

**[Figure 3]**

Expected returns on facilities result from netting expected default losses out of committed returns, that is, they factorize both committed returns and firms' probability of default. Hence, the finding that expected returns do not appear to compensate adequately for risk during the subprime crisis could be caused by the rise in corporate default probability in this period. Indeed, Figure 3 shows that average default probability intensely increased from 0.029 in 2007:Q4, at the breakout of the downturn, to a peak of 0.244 in 2008:Q4, that is, default probability grew more than eight times in a year.

In line with standard financial works on the risk–return trade-off and with previous studies suggesting that lenders are risk averse (Ratti 1980, Hughes et al. 1995, 1996, 1999, Hughes and Moon 1997, Angelini 2000), our base analysis uses expected returns to study whether the returns on credit lines play a risk-compensating role. However, to shed light on whether the increase in corporate default probability affected the risk-compensating role of returns during the subprime crisis, we also examine the relation between facilities' committed returns and corporate risk in this period. Specifically, without netting expected default losses out of committed returns, we repeat the analysis in which the empirical model includes an interaction term between risk and the crisis. Panel B of Table 6 reports the main results of this empirical exercise.[27]

The interaction terms are statistically nonsignificant across specifications in Panel B of Table 6. Hence, in contrast to the findings suggesting that the subprime crisis affects the risk-compensating role of expected returns, the downturn seems to have no effect on the relation between committed returns and risk. In addition, the coefficients of risk are significant and positive when the dependent variables are committed total and AISD returns (respectively columns (1) and (2) of Panel B), whereas this coefficient is not statistically significant in the specification corresponding to committed AISU returns (column (3) of Panel B). That is, risk seems to be traded off with committed total and AISD returns, but not with committed AISU returns. These outcomes regarding how risk relates to committed returns are similar to those in our base analysis, when we use expected returns as dependent variables and do not include interaction terms between risk and the crisis. Thus, since the difference between expected and committed returns is whether expected default losses are netted out or not, the results in Panel B support the conclusion that the increase in the corporate default probability during the subprime crisis seems to help explain why expected returns insufficiently compensate for risk in this period.

---

[27] Complete results are available upon request.

**6. Increases in risk and returns**

To enrich the study of the relation between risk and return in the market for credit lines, we use probit regression models to examine whether increases in risk between the previous and current quarters modify the probability of expected returns increasing in the following quarter. Table 7 reports the main results of this analysis.[28] The dependent and key independent variables of the analysis are (0, 1) indicator variables. In Panel A, the dependent variables are equal to one in columns (1) to (3) if the expected total, AISD, and AISU returns, respectively, increase between the current quarter, quarter $t$, and quarter $t + 1$, and zero otherwise. The variable measuring increases in risk is equal to one if corporate risk increases between $t - 1$ and $t$, and zero otherwise. In Panel B, we focus on situations in which the increases in risk and expected returns are both large. Specifically, the variable proxying for increases in risk is equal to one in this second panel if the increase in risk between $t - 1$ and $t$ is in the upper quartile of the sample distribution of quarterly increases in risk. Similarly, the dependent variables of the three specifications in Panel B are equal to one if the increase between $t$ and $t + 1$ of the corresponding expected return is above the 75th percentile of the sample distribution of the quarterly increases in this return. Along with the remainder of the control variables in the base model of our analysis, the probit models include borrower, firm credit rating, quarter, and lead lender indicator variables. Again as in the base model, we estimate the models with standard errors robust to within-facility dependence and heteroscedasticity.

[Table 7]

In Panel A of Table 7, the coefficient of the variable measuring increases in risk is statistically significant and positive when expected AISD returns is the dependent variable. Therefore, increases in risk seem to augment the probability of observing increases in this type of returns.

---

[28] Complete results are available upon request.

Specifically, according to the average marginal effect, the probability that expected AISD returns increase grows by 11.7% after corporate risk rises.

The coefficient of the variable that captures increases in risk in the specification of expected AISU returns is not significant in statistical terms. This outcome is in line with our findings regarding the relation between levels of risk and this type of returns. Indeed, such a result provides additional evidence supporting the conclusion that lenders could be receiving insufficient compensation for having credit commitments available for quality-deteriorating borrowers.

In Panel B of Table 7, the coefficient of increases in risk is also significant and positive in the specification of expected AISD returns. In relation to the average marginal effect, it is higher than in Panel A; that is, the rise in the probability of observing a high increase in expected AISD returns after a high increase in risk is greater than if we focus on plain increases. In particular, a high increase in risk (i.e., an increase lying in the upper quartile of the distribution of quarterly increases in risk) augments by 12.8% the probability of observing a high increase in expected AISD returns (i.e., an increase over the 75th percentile of the distribution of the quarterly increases in this kind of returns). We obtain a similar outcome for total expected coupon returns. In this case, the average marginal effect equals 8.7%. Nevertheless, since the coefficient of increases in risk is not significant in column (1) of Panel A, increases in risk and the probability that total expected returns rise seem to have a heterogeneous relation, with this probability being affected only when increases in returns and risk are high.

As in Panel A of Table 7, the coefficient of increases in risk is not significant in the specification of Panel B corresponding to expected AISU returns. Thus, consistent with our previous results regarding this type of returns, high increases in risk do not seem to augment the probability of obtaining large compensations from AISU returns either.

## 7. Conclusion

The credit risk intrinsic to the lending relationship is qualitatively different in credit lines. Since the borrower can draw down on them at contractually set terms, revolving facilities protect the firm against having to face worse credit conditions due to a deterioration of corporate quality (Berg et al. 2016). This paper analyzes whether returns help protect the other contractual party, that is, whether returns on revolving loan agreements compensate the lender for increases in borrower risk. If this were the case, we would observe a trade-off between risk and return; that is, all else being equal, increases in risk would be associated with a greater probability of obtaining a higher yield.

Despite credit lines being a dominant form of corporate financing, only Asarnow and Marker (1995) have studied whether risk and return are positively related in the market for credit lines. We follow the path their classical work has opened. Specifically, besides performing a univariate analysis similar to theirs, we provide the first multivariate analysis of how borrower risk relates to lenders' expected returns on revolving facilities. Our research is highly demanding in terms of data. Indeed, we use data from five commercial databases, two linking databases, and a manually gathered dataset providing information not covered or insufficiently covered by commercial databases.

Overall, in contrast to Asarnow and Marker's (1995), our findings support the hypothesis of a positive relation between corporate risk and facilities' expected returns. Bearing in mind that the relation of returns to interest rates and fees is not necessarily straightforward, this result is consistent with those noting that spreads and fees are increasing in corporate risk (Strahan 1999, Asquith et al. 2005). We also observe that increases in risk augment the probability of observing increases in expected AISD returns. A similar effect is found for expected total and AISD returns when we focus on large increases in both risk and these types of returns. Nevertheless, two kinds of mispricing also appear to be present in the market for corporate facilities. First, no risk premium seems to exist when we analyze the relation between risk and the expected AISU return. This outcome, also observed by Asarnow and Marker (1995), implies that the market for credit lines could be underpricing the risk of deteriorating borrowers increasing their drawdowns on lines of credit. Second, in line with the instability that seems to characterize the risk–return trade-off in stock markets through the business cycle (Ghysels et al. 2014, 2016), the relation between risk and return reversed in the subprime crisis, with expected returns decreasing in risk during this critical time. Regarding this reversion, the sharp rise in expected default losses during the recession seems to have eroded the risk-compensating role of expected returns.

# Appendix A. Variable definitions

| | | |
|---|---|---|
| **Table A1** | | |
| This table presents the definitions of the variables used in our analysis. Following Roberts and Sufi (2009), to calculate the variables grouped as firm characteristics, except for equity volatility, we use quarterly accounting values and then average the results from quarter $t$ to quarter $t-3$. Variable changes are calculated relative to the previous quarter. Variables from Compustat are in bold. Data on firms' stock returns are from the CRSP. Data on facilities were manually collected or are from DealScan. | | |
| Return | Return (conditional on no default) | (Income from spread charged on drawdowns and from annual, commitment, utilization and upfront fees)/(Outstanding borrowings + 0.5 * Unused portion of the commitment) |
| | Probability of default (PD) | Following Bharath and Shumway (2008), N(-DD), with N being the cumulative standard normal distribution function and DD = $[\ln((E + F)/F) + r - 0.5 * \sigma^2]/\sigma$, where (i) E = **prccq** * **cshoq**, (ii) F = **dlcq** + 0.5 * **dlttq**, (iii) r is the firm's annual stock return, calculated by cumulating monthly stock returns over the previous 12 months, and (iv) $\sigma$ = (E/(E + F)) * $\sigma_E$ + (F/(E + F)) * (0.05 + 0.25 * $\sigma_E$), where $\sigma_E$ is the annualized percent standard deviation of returns, calculated from monthly stock returns over the previous 12 months.[a] |
| | Expected return | (1 - PD) * Return + PD * Return * 0.348 |
| | AISD return (conditional on no default) | (Income from spread charged on drawdowns and from annual and utilization fees)/(Outstanding borrowings + 0.5 * Unused portion of the commitment) |
| | Expected AISD return | (1 - PD) * AISD return + PD * AISD return * 0.348 |
| | AISU return (conditional on no default) | (Income from annual, commitment and upfront fees)/(Outstanding borrowings + 0.5 * Unused portion of the commitment) |
| | Expected AISU return | (1 - PD) * AISU return + PD * AISU return * 0.348 |
| All-in spreads | AISD | Spread over LIBOR loans[b] + Annual fee + Utilization fee |
| | AISU | Commitment fee + Annual fee |
| Firm characteristics | Equity volatility | 12-month standard deviation of the firm's daily stock return |
| | Leverage | 1/((**dlcq** + **dlttq**)/**oibdpq**) |
| | Coverage | **oibdpq/xintq** |
| | Capital expenditures | **capxq/atq** |
| | Net worth | **atq – ltq** |
| | Current ratio | **actq/lctq** |
| | Profitability | **oibdpq/atq** |
| | Size | Log of **atq** |
| | Market-to-book ratio | (**atq** - (**atq** - **ltq** - **pstkl** + **txditcq**) + (**prccq** * **cshoq**))/**atq** |
| | Tangibility | **ppentq/atq** |

---

[a] If the firm has no outstanding debt, its probability of default is set to zero.
[b] Or spread over ABR loans, if available, and the credit line does not allow LIBOR loans.

| | | |
|---|---|---|
| | Kaplan–Zingales index | -1.001909 * ((**ibq** + **dpq**)/lagged **ppentq**) + 0.2826389 * (**atq** - **ceqq** - **txditcq** + (**prccq** * **cshoq**))/**atq** + 3.139193 * ((**dlcq** + **dlttq**)/(**dlcq** + **dlttq** + **seqq**)) - 39.3678 * (four-quarter moving average of **dvq**/lagged **ppentq**) - 1.314759 * (**cheq**/lagged **ppentq**) |
| | Monitoring cost | ((Δ**actq** – Δ**cheq**) – (Δ**lctq** – Δ**dlcq** - Δ**txpq**) - **dpq**)/**atq** |
| | Z-Score | 1.2 * ((**actq** - **lctq**)/**atq**) + 1.4 * (**req**/**atq**) + 3.3 * (**piq**/**atq**) + 0.6 * ((**prccq** * **cshoq**)/**ltq**) + 0.999 * (**saleq**/**atq**) |
| Facility characteristics | Secured | 1 if the facility is secured and 0 otherwise |
| | Syndicated | 1 if the credit line is provided by a syndicate of lenders and 0 otherwise |
| | Maturity | Log of the facility maturity specified in the credit contract |
| | Amount | Log of the amount committed under the facility |
| | Purpose | 1 if the facility can be used for corporate restructuring, that is, leveraged buyouts, mergers and acquisitions, and stock repurchases, and 0 otherwise |
| | Technical default | 1 if the financial covenants of the facility are breached by the borrower and 0 otherwise |
| | Nonzero outst. borrowings | 1 if the facility has positive outstanding borrowings and 0 otherwise |
| Fees | Annual fee | 1 if the facility includes a fee charged on the entire commitment amount, regardless of usage, and 0 otherwise |
| | Commitment fee | 1 if the facility includes a fee charged on the unused amount of the commitment and 0 otherwise |
| | Utilization fee | 1 if the facility includes a fee charged on the drawn amounts if and while a usage threshold is exceeded and 0 otherwise |
| | Upfront fee | 1 if the facility includes a single charge fee paid at origination and 0 otherwise |
| Other | Crisis | 1 if the quarter is in the period of the subprime crisis (according to NBER, 2007:Q4–2009:Q1) lagged by one quarter and 0 otherwise |
| | Lenders' systematic risk | Estimated $\beta$ coefficient resulting from regressing lenders' excess returns on market excess returns[c] |

---

[c] The risk-free rate used to compute excess returns is the three-month Treasury bill rate, as reported by the Federal Reserve Bank of St. Louis. Market excess returns are computed under the assumption that the market portfolio is given by the S&P 500 index.

**Appendix B. Characteristics of interest rate and fee schemes**

This appendix discusses the main features of the interest rates and fees of credit lines. We first analyze base rates and then proceed to describe spreads. Finally, we will focus on fees.

Under most LIBOR loans, the base rate is the prevailing LIBOR for a term equivalent to the maturity of the drawdown, as decided by the borrower among those offered by the credit agreement, typically one, two, three, or six months. However, a marginal proportion of credit agreements fix the maturity period and, hence, the applicable LIBOR. In our sample, 1.23% and 4.94% of the credit lines fix a three- and a one-month maturity, respectively.[d]

Some contracts establish the base rate ultimately used to determine the interest rate on a LIBOR loan to be the greater of the applicable LIBOR or a fixed percentage (2.67% of the sample lines). Additionally, for very few credit agreements, the final interest rate on LIBOR loans is not the sum of the applicable LIBOR plus a margin but, instead, the greater value between this sum and a fixed percentage (0.41%).

Regarding ABR loans, the base rate is a given rate or the greatest among a set of rates to which a percentage is sometimes added.[e] In all sample lines where no set of rates is competing to be the ABR (13.01% of the sample lines with ABR), the latter is always a prime rate, either the prime rate reported by *The Wall Street Journal* or one of the lenders' official prime rates. When multiple rates compete to be the ABR, a prime rate is always among them. Another candidate present in most credit agreements with an ABR is the federal funds rate plus a margin (85.71%), which is commonly equal to 50 basis points (bps). Some contracts include among the competing candidates a fixed percentage (1.92%). An increasing tendency during the sample period is to include the one- or three-month LIBOR plus a percentage as one of the candidates. In this sense, just 3.44% of the sample credit lines with an ABR that were active before the end of the 2007 crisis included the LIBOR among the competing rates, whereas this percentage rises to 71.81% among credit lines with an ABR outstanding after 2009:Q2. This sharp increase can be framed as a general tendency to make facility pricing more flexible and adaptable to changing circumstances.

---

[d] In addition, to take into account reserve requirements for Eurocurrency liabilities and, thus, compensate lenders for additional costs associated with obtaining funds from the Eurodollar market, the LIBOR is transformed into what credit contracts usually call the "adjusted LIBOR." We can, however, disregard the difference between the LIBOR and the adjusted LIBOR in our analysis of the risk–return relation, since those requirements are zero during the sample period.

[e] For instance, according to the fifth amendment dated November 12, 2008, to the amended and restated credit agreement dated November 4, 2005, of Petroleum Development Corporation, the base rate of ABR loans is a "rate per annum equal to the greatest of (a) the Prime Rate in effect on such day, (b) the Base CD Rate in effect on such day plus 1%, (c) the Federal Funds Effective Rate in effect on such day plus ½ of 1%, and (d) the Adjusted LIBO Rate for a one month Interest Period on such day … plus 1%."

Margins added to the applicable base rates may be fixed or variable. If they are fixed, the spread charged on drawdowns is constant for the life of the contract. Among samples lines that allow borrowing ABR (LIBOR) loans, 39.44% (18.98%) charge fixed spreads on this type of loans. Variable margins change over time according to an agreed schedule or are dependent on one or more pricing criteria that reflect corporate performance or credit line usage (e.g., Appendix A of Asquith et al. 2005). Most credit lines determine applicable spreads in terms of a single pricing criterion, but multiple criteria are not unusual (see Table C2 in Appendix C).

If a firm breaches a credit line's financial covenants—that is, is in technical default—and does not obtain a waiver of compliance on this violation, the interest rate in effect is increased by a default margin. Additionally, some credit agreements that give borrowers the chance to choose between ABR and LIBOR loans restrict this option under a technical default, allowing only ABR loans. For sample credit lines involved in technical defaults, the average default margin to be added to the applicable interest rate is 243.06 bps and 22.22% of those credit lines restrict the types of available loans. However, technical default rarely leads to margin increases: Such penalizations are waived for most covenant violations (83.33%).

Commitment fees are charged on the unused parts of credit lines, whereas annual fees are levied on the entire commitment amounts, used or not. Utilization fees are charged on used amounts if and while an agreed upon usage threshold is exceeded.[f] Among sample lines, 79.01%, 17.49%, and 5.14% have commitment, annual, and utilization fees, respectively. Just as interest rate spreads, fees can be constant or variable. In the latter case, one or more pricing criteria determine the applicable fee. The percentage of credit lines with fixed commitment (annual, utilization) fees over the total number of sample lines with this type of fee is 40.62% (9.41%, 60.00%).

---

[f] In some cases, DealScan confuses utilization fees with situations in which the interest rate spread depends on facility usage. This typically happens when a credit line (1) offers the option to borrow LIBOR and ABR loans, (2) the spread of ABR loans is fixed, and (3) the spread of LIBOR loans depends on a set of criteria, one of them being usage. An example is the amended and restated agreement, dated March 31, 2006, between Allergan Inc. and a syndicate of financial institutions. This agreement provides the option to borrow ABR loans with a fixed spread equal to zero. It also allows Allergan Inc. to borrow LIBOR loans. The spread on the latter depends on the firm's rating and whether the usage-to-principal ratio is above or below 50%. Therefore, in contrast to the definition of a utilization fee, a level of usage over 50% implies an extra charge on LIBOR loans, but not on the entire amount used. Nevertheless, DealScan considers that this credit line includes a utilization fee.

# Appendix C. Pricing criteria

**Table C1**

This table presents the pricing criteria that, according to credit line contracts, are used to calculate variable spreads and fees. The column *Type* broadly classifies the pricing criteria. Based on this classification, the column *Id.* provides an identification of each pricing criterion. The column *Definition* describes how pricing criteria are calculated in terms of variables from Compustat (in bold), DealScan (in italics), Capital IQ (underlined), and Datastream (in underlined italics), or variables based on manually gathered data (plain text). The column *Credit lines* indicates the number and percentage (over the total number of facilities with pricing grids) of the sample lines per pricing criterion. The variables based on manually gathered data are defined as follows: borr, outstanding borrowings on a credit line; borrbase, the dollar value of a credit line's borrowing base; lc, the dollar value of outstanding letters of credit under a credit line; lecrq, the dollar value of a firm's total amount of outstanding letters of credit; and unusedav, a credit line's unused available amount.

Technical notes: (1) If the SEC filings do not mention a credit line's unused available amount for a quarter, we calculate it as borrbase – borr – lc or, if the credit line does not have a program to support the issuance of letters of credit, as borrbase – borr. If the credit line does not have a borrowing base or borrbase is missing, we use *facilityamt* – borr – lc or *facilityamt* – borr. (2) If the SEC filings do not provide information about letters of credit outstanding under a facility but this information is required to calculate a pricing criterion, we assume that lc is zero. (3) The income statement variables are measured on a four-quarter rolling basis. (4) To obtain the quarterly values of firms' rent expenses, we generate the variable **xrentq**, which is equal to Compustat's annual variable **xrent** divided by four. (5) Following Demerjian and Owens (2016), we use four-quarter lagged debt in current liabilities (**dlcq**$_{t-4}$) as a proxy of senior debt payments during the past year. (6) Quarterly values for capital expenditures (**capxq**) and dividend payments (**dvq**) are obtained from the year-to-date cash flow statement data. (7) If the Capital IQ variable for senior debt (totsrdbt) is missing, we assume that the Compustat annual variable for subordinated debt (**ds**) remains constant during the fiscal year and we calculate quarterly senior debt as the difference between total debt and the latter variable (**dlcq + dlttq - ds**). We do not use the Capital IQ variable for subordinated debt (totsubdbt) due to the large number of missing values. If senior secured debt (totsrsecureddbt) or secured debt (secureddbt) is required to calculate a pricing criterion but is missing from the Capital IQ data, we use the Compustat variable for mortgage and other secured debt (**dm**).

| Type | Id. | Definition | Credit lines | |
|---|---|---|---|---|
| | | | Number | Percent. |
| Availability | A1 | unusedav | 22 | 5.55% |
| Adjusted availability | B1 | unusedav+**cheq** | 3 | 0.75% |
| | B2 | unusedav-2.5 | 1 | 0.25% |
| | B3 | unusedav-3.5 | 1 | 0.25% |
| Availability-to-borrowing base ratio | C1 | unusedav/borrbase | 1 | 0.25% |
| Availability-to-facility amount ratio | D1 | unusedav/*facilityamt* | 1 | 0.25% |
| Usage | E1 | borr+lc | 3 | 0.75% |
| | E2 | borr | 2 | 0.50% |
| Usage-to-borrowing base ratio | F1 | (borr+lc)/borrbase | 17 | 4.26% |
| Usage-to-facility amount ratio | G1 | (borr+lc)/*facilityamt* | 12 | 3.01% |
| | G2 | borr/*facilityamt* | 1 | 0.25% |
| Debt-to-EBITDA ratio | H1 | (**dlcq+dlttq**)/**oibdpq** | 175 | 43.86% |
| Adjusted debt-to-EBITDA ratio | I1 | (**dlcq+dlttq-chq**)/**oibdpq** | 2 | 0.50% |
| | I2 | (**dlcq+dlttq-cheq**)/**oibdpq** | 2 | 0.50% |
| | I3 | **(dlcq+dlttq**-lecredq-**chq)/oibdpq** | 3 | 0.75% |
| | I4 | **(dlcq+dlttq**+lecredq-**cheq)/oibdpq** | 2 | 0.50% |
| | I5 | **(dlcq+dlttq**-(**chq**-25)**)/oibdpq** | 3 | 0.75% |
| | I6 | **(dlcq+dlttq**-(**cheq**-30)**)/oibdpq** | 2 | 0.50% |
| | I7 | **(dlcq+dlttq**-(**cheq**-100)**)/oibdpq** | 1 | 0.25% |
| | I8 | **(dlcq+dlttq**-min{5,**chq**})**/oibdpq** | 1 | 0.25% |
| | I9 | **(dlcq+dlttq**-min{80,**chq**})**/oibdpq** | 3 | 0.75% |
| | I10 | **(dlcq+dlttq**-min{35,**cheq**})**/oibdpq** | 2 | 0.50% |
| | I11 | **(dlcq+dlttq**-min{100,**cheq**})**/oibdpq** | 1 | 0.25% |
| | I12 | **(dlcq+dlttq**-min{150,**cheq**-50})**/oibdpq** | 7 | 1.75% |
| Debt-to-capitalization ratio | J1 | **(dlcq+dlttq)/(dlcq+dlttq+atq-intanq-ltq)** | 5 | 1.25% |
| | J2 | **(dlcq+dlttq)/(atq+dlcq+dlttq -ltq)** | 4 | 0.75% |
| | J3 | **(dlcq+dlttq**-max{0,**chq**-10})**/(atq-intanq-ltq)** | 2 | 0.50% |
| | J4 | **(dlcq+dlttq**-max{0,**cheq**-10})**/(atq-intanq-ltq)** | 1 | 0.25% |
| Senior debt-to-EBITDA ratio | K1 | totsrdbt**/oibdpq** | 11 | 2.76% |
| Senior debt-to-capitalization ratio | L1 | totsrdbt**/(atq-intanq-ltq+(dlcq+dlttq**-totsrdbt**))** | 1 | 0.25% |
| Senior secured debt-to-EBITDA ratio | M1 | totsrsecureddbt**/oibdpq** | 2 | 0.50% |
| Adjusted senior secured debt-to-EBITDA ratio | N1 | (totsrsecureddbt-**cheq**)**/oibdpq** | 1 | 0.25% |
| Adjusted secured debt-to-EBITDA ratio | O1 | (secureddbt-borr+((borr+borr$_{t-1}$+borr$_{t-2}$+borr$_{t-3}$)/4))**/oibdpq** | 3 | 0.75% |
| Adjusted debt-to-EBITDAR ratio | P1 | (8***xrentq**+lecrq+**dlcq+dlttq)/(oibdpq+xrentq)** | 2 | 0.50% |
| | P2 | (8***xrentq**+**dlcq+dlttq)/(oibdpq+xrentq)** | 2 | 0.50% |
| | P3 | (7***xrentq**+**dlcq+dlttq)/(oibdpq+xrentq)** | 1 | 0.25% |
| | P4 | (7***xrentq**+lecrq+**dlcq+dlttq)/(oibdpq+xrentq)** | 5 | 1.25% |
| Liabilities-to-capitalization ratio | Q1 | **ltq/(atq-intanq-ltq)** | 2 | 0.50% |
| Adjusted liabilities-to-EBITDA ratio | R1 | **(ltq-apq-xacc-drltq)/(oibdpq)** | 1 | 0.25% |
| Fixed charge coverage ratio | S1 | **(piq-capxq)/dlcq**$_{t-4}$ | 2 | 0.50% |
| | S2 | **(oibdpq+xrentq-capxq)/(xrentq+xintq)** | 1 | 0.25% |
| | S3 | **(oibdpq+xrentq)/(xrentq+xintq)** | 1 | 0.25% |
| | S4 | **(piq-capxq-dvq)/(dlcq**$_{t-4}$+**xintq)** | 1 | 0.25% |
| | S5 | **oibdpq/xintq** | 1 | 0.25% |
| | S6 | **(oibdpq+xrentq-txtq-capxq-dvq)/(dlcq**-borr+**xrentq+xintq)** | 1 | 0.25% |
| | S7 | **(oibdpq+xrentq)/(dlcq**$_{t-4}$+**xrentq+xintq)** | 3 | 0.75% |
| EBITDA | T1 | **oibdpq** | 4 | 1.00% |
| Equity-to-assets ratio | U1 | **seqq/atq** | 1 | 0.25% |
| Rating | V1 | End-of-quarter **splticrm** | 90 | 22.56% |
| Credit default swap spread | W1 | *xr 5y $ - cds prem. mid* | 1 | 0.25% |

| Time elapsed since origination | X1 | Time elapsed since origination | 4 | 1.00% |

**Table C2. Credit lines with spreads or fees defined by more than one criterion**

This table describes situations in which more than a single criterion determines the applicable spreads or fees. The criteria are identified according to column *Id.* in Table C1.

| Criteria | Credit lines | |
| --- | --- | --- |
| | Number | Percentage |
| E1 & V1 | 1 | 0.25% |
| G1 & V1 | 6 | 1.50% |
| G2 & V1 | 1 | 0.25% |
| H1 & V1 | 6 | 1.50% |
| W1 & V1 | 1 | 0.25% |
| A1, E2 & X1 | 4 | 1.00% |
| A1, T1 & X1 | 1 | 0.25% |

# Appendix D. Credit line purposes according to credit line contract

## Table D1

This table presents the purposes for which borrowings from sample facilities can be used according to credit contracts. The columns show the possible purposes. An X indicates that the type of credit line allows borrowers to use drawdowns for the purpose denoted in the corresponding column. The last column shows the number of sample credit lines by type. Support for letters of credit and payments of fees and expenses associated with credit lines are not included as purposes. Cash management exposure is included in working capital or corporate purposes. The data were obtained from the credit contracts of sample credit lines.

| Types | Work. cap. or corp. purposes | Debt repaym. | Mergers and acquis. | Capital expend. | Stock buyback | Specific event | Joint ventures | Specific event | Purch. convert. notes | Foreign exch. | Total (over 496) |
|---|---|---|---|---|---|---|---|---|---|---|---|
| 1 | X | | | | | | | | | | 86 (17.70%) |
| 2 | X | X | | | | | | | | | 81 (16.67%) |
| 3 | X | X | X | | | | | | | | 68 (13.99%) |
| 4 | X | X | X | X | | | | | | | 31 (6.38%) |
| 5 | X | X | X | X | X | | | | | | 6 (1.23%) |
| 6 | X | X | X | X | | X | | | | | 11 (2.26%) |
| 7 | X | X | X | | X | | | | | | 9 (1.85%) |
| 8 | X | X | | X | | | | | | | 35 (7.20%) |
| 9 | X | X | | X | X | | | | | | 3 (0.62%) |
| 10 | X | X | | X | X | | X | | | | 3 (0.62%) |
| 11 | X | X | | | X | | | | | | 5 (1.03%) |
| 12 | X | X | | | | X | | | | | 8 (1.65%) |
| 13 | X | | X | | | | | | | | 45 (9.26%) |
| 14 | X | | X | X | | | | | | | 34 (7.00%) |
| 15 | X | | X | X | X | | | | | | 4 (0.82%) |
| 16 | X | | X | | X | | | | | | 5 (1.03%) |
| 17 | X | | X | | | | X | | | | 3 (0.62%) |
| 18 | X | | X | | X | | X | | | | 1 (0.21%) |
| 19 | X | | | X | | | | | | | 24 (4.94%) |
| 20 | X | | | X | X | | | | | | 3 (0.62%) |
| 21 | X | | | X | X | | | | | | 2 (0.41%) |
| 22 | X | | | | | | | X | | | 3 (0.62%) |
| 23 | X | | | | | | | | X | | 2 (0.41%) |

| 24 | X |  |  |  |  |  |  |  |  | X | 2 (0.41%) |
| 25 |  |  | X | X |  |  |  |  |  |  | 1 (0.21%) |

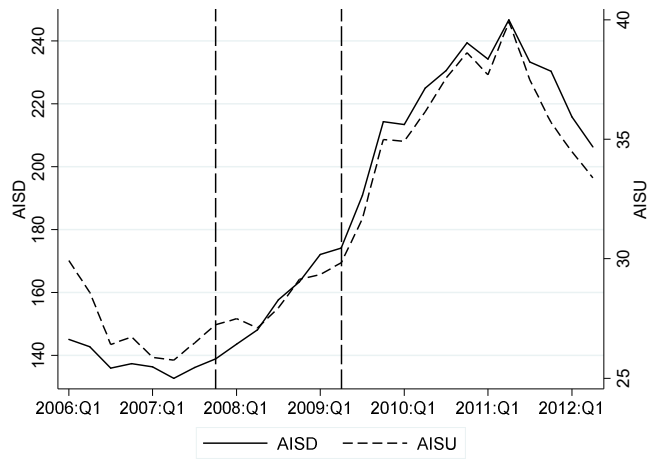

**Figure 1. AISD and AISU.** This figure represents the sample average AISD and AISU by quarter. The AISD and AISU are measured on the left- and right-hand axes, respectively. The vertical dashed lines mark the start (2007:Q4) and end (2009:Q2) of the 2007 crisis according to the NBER. The AISD and AISU are defined in Appendix A.

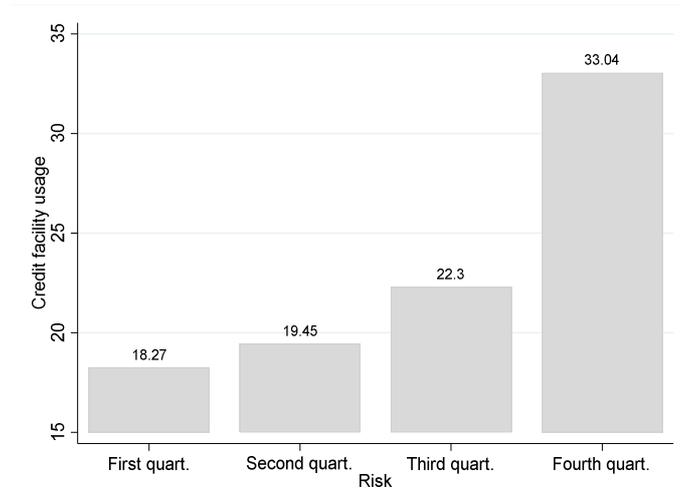

**Figure 2. Risk and credit facility usage.** This figure represents the average usage-to-commitment ratio (in percentages) of the sample credit lines by quartiles of corporate risk.

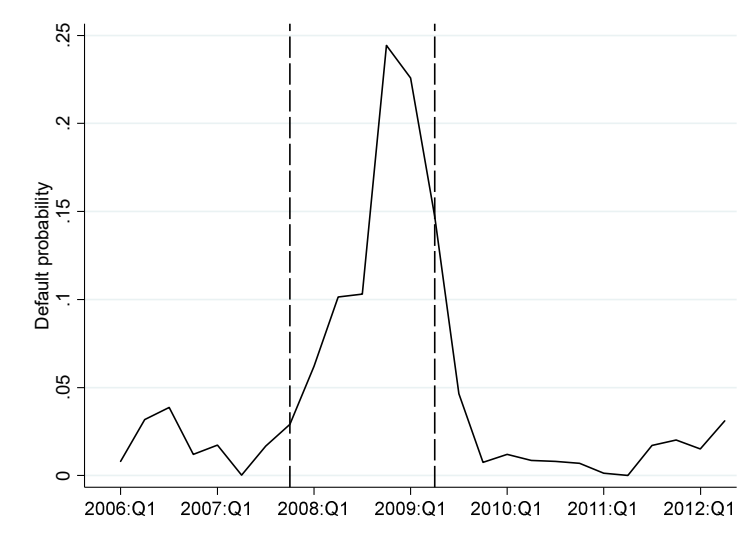

**Figure 3. Corporate default probability.** This figure represents the average default probability of sample firms by quarter. The vertical dashed lines mark the start (2007:Q4) and end (2009:Q2) of the subprime crisis according to the NBER. The corporate default probability is defined in Appendix A.

**Table 1. Summary statistics: Revolving facility characteristics**

This table presents summary statistics—means, standard errors (SE), and medians—on the facility characteristics for two datasets. The first (sample) is the dataset on which our analysis is based. The second dataset (DealScan) consists of all the revolving lines of credit in DealScan outstanding in the sample period, denominated in US dollars, and provided to non-financial US firms. Statistics on the purpose of the facilities in DealScan are not provided due to substantial differences between our variable and that of DealScan (see Appendix D). To calculate the statistics referring to our sample, we proceed as if there were one observation per facility. However, the statistics of technical default and nonzero outstanding borrowings are calculated over firms and the total number of observations, respectively. The variables are defined in Appendix A.

|  | Sample | | | DealScan | | |
|---|---|---|---|---|---|---|
|  | Mean | SE | Median | Mean | SE | Median |
| Secured line (1, 0) | 0.67 | 0.47 | 1 | 0.79 | 0.41 | 1 |
| Syndicated line (1, 0) | 0.90 | 0.30 | 1 | 0.98 | 0.12 | 1 |
| Maturity (months) | 44.22 | 16.94 | 46 | 50.08 | 18.72 | 60 |
| Amount ($mil.) | 266.03 | 365.81 | 125 | 266.09 | 596.79 | 90 |
| Purpose (1, 0) | 0.50 | 0.50 | 0 | - | | |
| Technical default | 0.17 | 0.38 | 0 | - | | |
| Nonzero outst. borrs. (1, 0) | 0.52 | 0.50 | 1 | - | | |
| Credit lines | 486 | | | 23,148 | | |
| Firms | 122 | | | 10,191 | | |

| | Sample | | | Compustat | | |
|---|---|---|---|---|---|---|
| | Mean | SE | Median | Mean | SE | Median |
| Risk | 0.03 | 0.02 | 0.03 | - | | |
| Leverage | 1.35 | 12.02 | 0.11 | 0.76 | 136.12 | 0.07 |
| Coverage | 152.09 | 1,557 | 7.71 | 36.81 | 817.19 | 4.23 |
| Capital expend. | 0.01 | 0.02 | 0.01 | 0.01 | 0.04 | 0.01 |
| Net worth | 882.31 | 1,381 | 320.26 | 1,081 | 5,461 | 98.09 |
| Current ratio | 1.92 | 1.05 | 1.68 | 4.19 | 49.53 | 1.85 |
| Profitability | 0.03 | 0.03 | 0.03 | -0.69 | 49.80 | 0.02 |
| Size ($ mil.) | 2,242 | 3,394 | 766.84 | 3,033 | 16,672 | 231.05 |
| Market to book ratio | 1.70 | 1.02 | 1.36 | 32.45 | 928.12 | 1.64 |
| Tangibility | 0.29 | 0.24 | 0.22 | 0.26 | 0.25 | 0.16 |
| Financial constraint | -0.26 | 11.45 | 0.73 | 315.60 | 75,487 | -0.07 |
| Monitoring cost | -0.01 | 0.02 | -0.01 | 0.16 | 44.46 | -0.01 |
| Line–quarter obs. | | 2,073 | | | 133,037 | |
| Firms | | 122 | | | 7,114 | |

**Table 2. Summary statistics: Firm characteristics**

This table presents summary statistics—means, standard errors (SE), and medians—on the firm characteristics for two datasets. The first (sample) is the dataset on which our analysis is based. The second dataset (Compustat) consists of all firm–quarter observations from non-financial US firms appearing in Compustat in the sample period. The variables are defined in Appendix A.

**Table 3. Univariate analysis**

This table presents the different types of mean expected annual returns of the credit lines for the whole sample and by risk category. The risk categories are the quintiles of the 12-month standard deviation of the sample firms' daily stock returns. To compute the mean annual return for either the whole sample or each risk category, we calculate, first, the average expected returns of the sample facilities per quarter; second, the compound cumulative return; and, third, the geometric average of this cumulative return. The types of returns are defined in Appendix A.

| Type | Quintiles of the st. dev. of daily stock returns | | | | | Total |
|---|---|---|---|---|---|---|
| | First | Second | Third | Fourth | Fifth | |
| Expected return (1) | 0.79% | 0.87% | 0.97% | 1.28% | 1.67% | 1.05% |
| Exp. AISD return (2) | 0.48% | 0.49% | 0.54% | 0.69% | 1.13% | 0.62% |
| Exp. AISU return (3) | 0.41% | 0.47% | 0.47% | 0.65% | 0.55% | 0.49% |

**Table 4. Risk-compensating role of return**

This table presents the results from regressions that analyze the risk–return relation in the market for credit lines. The dependent variable in both columns (1) and (2) is the expected total coupon return and that in columns (3) and (4) is the expected AISD and AISU returns, respectively. Along with fixed effects, column (1) only includes risk as the regressor. The specification in column (2) is that of the base model. Risk is measured by the 12-month standard deviation of sample firms' daily stock returns. The regressions include borrower, firm credit rating, quarter, and lead lender indicator variables. Statistical significance at the 10%, 5%, and 1% levels is denoted by *, **, and **, respectively. All the regressions are conducted with standard errors robust to within-facility dependence and heteroscedasticity. The variables are defined in Appendix A.

|  | Fixed effects (1) | Base model (2) | Exp. AISD return (3) | Exp. AISU return (4) |
|---|---|---|---|---|
| Intercept | 0.251 (0.528) | 2.278 (1.473) | 1.005 (1.096) | 1.234 (1.175) |
| Risk | 11.100*** (2.954) | 7.648*** (2.754) | 9.857*** (2.371) | -1.837 (2.412) |
| Secured | - | 0.304** (0.119) | 0.154 (0.110) | 0.137* (0.078) |
| Syndicated | - | 0.325* (0.187) | 0.378* (0.215) | -0.046 (0.119) |
| Maturity | - | -0.617*** (0.181) | -0.313*** (0.076) | -0.299 (0.188) |
| Amount | - | -0.052 (0.064) | 0.004 (0.054) | -0.072 (0.050) |
| Purpose | - | -0.106 (0.117) | -0.166 (0.115) | 0.029 (0.067) |
| Annual fee | - | 1.730*** (0.426) | 1.121*** (0.309) | 1.064*** (0.325) |
| Commitment fee | - | 1.708*** (0.310) | 0.748*** (0.194) | 1.002*** (0.282) |
| Utilization fee | - | 0.186 (0.134) | -0.318*** (0.105) | 0.091 (0.114) |
| Upfront fee | - | 0.171 (0.128) | 0.155 (0.098) | 0.048 (0.095) |
| Technical default | - | -0.198 (0.161) | -0.125 (0.181) | -0.124 (0.082) |
| Nonzero outst. borrow. | - | 0.411*** (0.078) | 0.773*** (0.064) | -0.394*** (0.073) |
| Leverage | - | -0.002 (0.001) | -0.001 (0.001) | -0.001** (0.000) |
| Coverage | - | 0.000 (0.000) | -0.000 (0.000) | 0.000 (0.000) |
| Capital expenditures | - | 6.671* (3.700) | 7.683** (3.562) | -1.137 (0.878) |
| Net worth | - | 0.000<br>(0.000) | 0.000<br>(0.000) | -0.000<br>(0.000) |
| Current ratio | - | -0.308***<br>(0.067) | -0.346***<br>(0.070) | 0.031<br>(0.033) |
| Profitability | - | -4.786***<br>(1.619) | -5.010***<br>(1.368) | 0.115<br>(0.848) |
| Size | - | -0.146<br>(0.179) | 0.109<br>(0.147) | -0.273**<br>(0.117) |
| Market-to-book ratio | - | 0.045<br>(0.062) | 0.014<br>(0.055) | 0.029<br>(0.031) |
| Tangibility | - | -1.125*<br>(0.671) | -2.079***<br>(0.536) | 1.049***<br>(0.379) |
| Financial constraint | - | -0.002<br>(0.002) | 0.000<br>(0.001) | -0.003<br>(0.002) |
| Monitoring cost | - | 1.010<br>(1.385) | -0.285<br>(1.145) | 1.649**<br>(0.800) |
| Crisis | - | -0.396***<br>(0.119) | -0.210**<br>(0.105) | -0.206**<br>(0.084) |
| Observations | 1,830 | 1,286 | 1,286 | 1,286 |
| $R^2$ | 0.594 | 0.734 | 0.742 | 0.695 |

**Table 5. Robustness checks**

This table presents the main results of robustness checks of the hypothesis that return plays a risk-compensating role in the market for credit lines. The table only shows the estimated coefficients of the variable measuring risk and, in row (8), squared risk. The columns differ in the dependent variable used, that is, the expected total coupon AISD, and AISU returns. Regarding the differences between robustness checks and the base model, in row (1), risk is measured by Altman's Z-score. In row (2), we drop lender fixed effects from the base model and include lenders' beta as a regressor. Rows (3) and (4) use alternative methods to amortize upfront fees. In row (3), the amount amortized each quarter is equal to the upfront fee divided by the number of quarters between the quarter in which the contract (either the original contract or the amendment including the fee) is settled and the earliest quarter between the quarter corresponding to the maturity date of the contract and the quarter in which the loan path of the facility terminates. In row (4), upfront fees are amortized only while the credit contracts that include them are not amended or terminated. In row (5), we use 14 months as the maturity threshold that divides facilities into those that do not have to hold capital for the unused portion of the commitment and those that do. In row (6), no maturity threshold is used to compute the returns. In row (7), instead of computing firm-specific controls using quarterly accounting values and then averaging the results from quarter $t$ to quarter $t-3$, we annualize flow variables to calculate these control variables. The regression model in row (8) includes a risk-squared term. The last two rows only take into account firms whose mean standard deviation of stock returns is above the median of all sample firms. These means and medians are calculated for the sample and the 2007 crisis periods in rows (9) and (10), respectively. The regressions include borrower, firm credit rating, and quarter indicator variables. Except in row (2), they also include lead lender fixed effects. Statistical significance at the 10%, 5%, and 1% levels is denoted by *, **, and ***, respectively. All the regressions are conducted with standard errors robust to within-facility dependence and heteroscedasticity. The variables are defined in Appendix A.

| | | Exp. return | Exp. AISD return | Exp. AISU return |
|---|---|---|---|---|
| Z-Score (1) | | -0.227*** (0.059) | -0.191*** (0.051) | -0.041 (0.033) |
| Lender's systematic risk (2) | | 6.370** (2.763) | 8.689*** (2.266) | -1.951 (2.525) |
| Upfront fee I (3) | | 7.796*** (2.764) | 9.857*** (2.371) | -1.693 (2.443) |
| Upfront fee II (4) | | 9.330*** (2.570) | 9.857*** (2.371) | -1.278 (1.021) |
| Capital for unused portion I (5) | | 7.648*** (2.754) | 9.857*** (2.371) | -1.837 (2.412) |
| Capital for unused portion II (6) | | 7.648*** (2.753) | 9.865*** (2.375) | -1.843 (2.411) |
| Annualized flow variables (7) | | 10.250*** (3.076) | 13.030*** (2.838) | -2.291 (2.821) |
| Squared term (8) | Risk | 12.550* (6.742) | 13.620** (5.888) | -3.289 (4.810) |
| | Squared risk | -36.840 (41.810) | -28.260 (39.020) | 10.920 (37.850) |
| Riskiest firms I (9) | | 5.598* (2.918) | 7.165*** (2.272) | -1.063 (2.507) |
| Riskiest firms II (10) | | 6.188** (2.807) | 7.884*** (2.261) | -1.182 (2.532) |

**Table 6. The 2007–2009 Crisis**

This table presents the results from analyzing the effects of the subprime crisis on the risk–return relation in the market for credit lines. The table shows the estimated coefficients of risk, the crisis, and the interaction term between these two variables. The inclusion of this interaction term is the only difference between the specifications used to obtain the results in Panel A of this table and those in columns (2) to (4) of Table 4. Panels A and B differ in that the dependent variables of the latter's specifications are committed (instead of expected) returns. Statistical significance at the 10%, 5%, and 1% levels is denoted by *, **, and **, respectively. All the regressions are conducted with standard errors robust to within-facility dependence and heteroscedasticity. The variables are defined in Appendix A.

| | **Panel A. Expected returns** | | |
|---|---|---|---|
| | Exp. return (1) | Exp. AISD return (2) | Exp. AISU return (3) |
| Risk | 11.330*** (2.629) | 12.460*** (2.697) | -0.606 (2.499) |
| Crisis | 0.006 (0.141) | 0.074 (0.128) | -0.072 (0.097) |
| Risk * Crisis | -14.440*** (3.962) | -10.230*** (2.959) | -4.831* (2.853) |
| Observations | 1,286 | 1,286 | 1,286 |
| $R^2$ | 0.743 | 0.747 | 0.698 |
| | **Panel B. Committed returns** | | |
| | Com. return (1) | Com. AISD return (2) | Com. AISU return (3) |
| Risk | 12.560*** (2.385) | 13.230*** (2.734) | -0.271 (2.063) |
| Crisis | -0.402** (0.159) | -0.357** (0.152) | -0.059 (0.091) |
| Risk * Crisis | -2.500 (3.738) | 1.043 (3.260) | -3.735 (2.393) |
| Observations | 1,398 | 1,398 | 1,398 |
| $R^2$ | 0.748 | 0.729 | 0.711 |

| Table 7. Increases in risk and returns |
| --- |
| This table presents the results from analyzing whether increases in risk raise the probability of observing increases in expected returns. To perform the analysis, we use probit regression models. In columns (1) to (3) of Panel A, the dependent variables are equal to one, respectively, if expected total, AISD, and AISU returns increase between the current and next quarters, and zero otherwise. In this panel, the key regressor (i.e., increases in risk) is an indicator variable that equals one if risk increases between the previous and current quarters, and zero otherwise. In Panel B, the dependent variables in columns (1) to (3) are also binary variables equal to one if expected total, AISD, and AISU returns, respectively, increase substantially, particularly if the increases in these returns between the current and next quarters are in the fourth quartile of the sample distributions of their respective quarterly increases. Similarly, the variable measuring increases in risk equals one in Panel B if the increase in risk between the previous and current quarters is in the upper quartile of the sample distribution of quarterly increases in risk, and zero otherwise. The table displays the estimated coefficients, standard errors (in parentheses), and average marginal effects of the main regressor in the probit models. Along with the remainder of the control variables in the base model of our analysis, the regressions include borrower, firm credit rating, quarter, and lead lender indicator variables. Statistical significance at the 10%, 5%, and 1% levels is denoted by *, **, and ***, respectively. All the regressions are conducted with standard errors robust to within-facility dependence and heteroscedasticity. |

| | | Panel A. Increases in risk | | |
| --- | --- | --- | --- | --- |
| | | Exp. return (1) | Exp. AISD return (2) | Exp. AISU return (3) |
| Increases in risk | Coefficient (stand. error) | 0.241 (0.150) | 0.421*** (0.163) | 0.046 (0.149) |
| | Average marginal effect | 0.070 | 0.117*** | 0.013 |
| Prob(Y=1) | | 0.377 | 0.373 | 0.372 |
| Pseudo-$R^2$ | | 0.223 | 0.250 | 0.261 |
| Observations | | 852 | 726 | 873 |
| | | Panel B. High increases in risk | | |
| | | Exp. return (1) | Exp. AISD return (2) | Exp. AISU return (3) |
| Increases in risk | Coefficient (stand. error) | 0.313** (0.152) | 0.470*** (0.149) | 0.181 (0.153) |
| | Average marginal effect | 0.087** | 0.128*** | 0.046 |
| Prob(Y=1) | | 0.343 | 0.356 | 0.335 |
| Pseudo-$R^2$ | | 0.230 | 0.252 | 0.289 |
| Observations | | 756 | 724 | 778 |